\documentclass[10pt,journal,compsoc]{IEEEtran}
\usepackage{epsfig,endnotes}
\usepackage[cmex10]{amsmath}
\usepackage{algorithmic}
\usepackage{array}
\usepackage{fixltx2e}

\usepackage{mdwmath}
\usepackage{amsmath}
\usepackage{multirow}
\usepackage{mathtools}

\usepackage{mdwtab}
\usepackage[flushleft]{threeparttable}
\usepackage{stfloats}
\usepackage[ruled,vlined]{algorithm2e}

\usepackage{slashbox}
\usepackage{caption}
\usepackage{subcaption}

\usepackage{threeparttable}

\usepackage{cite}

\makeatletter
 \let\@copyrightspace\relax
\makeatother

\usepackage{epstopdf}

\usepackage{mathrsfs}
\usepackage[scr=rsfs,cal=boondox]{mathalfa}

\begin{document}
%don't want date printed
\date{}
%make title bold and 14 pt font (Latex default is non-bold, 16 pt)
\title{BubbleMap: Privilege Mapping for Behavior-based Implicit Authentication Systems}

\author{
%\vspace{-0.4cm}
\IEEEauthorblockN{Yingyuan Yang\IEEEauthorrefmark{1}, Xueli Huang\IEEEauthorrefmark{3}, Jiangnan Li\IEEEauthorrefmark{2}, and Jinyuan Stella Sun\IEEEauthorrefmark{2}\\}
\IEEEauthorblockA{\IEEEauthorrefmark{1}University of Illinois, Springfield, IL, 62629 USA Email: yyang260@uis.edu\\}
\IEEEauthorblockA{\IEEEauthorrefmark{3}Temple University, Philadelphia, PA, 19122 USA Email: tuc36161@temple.edu\\}
\IEEEauthorblockA{\IEEEauthorrefmark{2}University of Tennessee, Knoxville, TN, 37996 USA Email: \{jli103, jysun\}@utk.edu
%\\Email: yyang260@uis.edu,tuc36161@temple.edu,jli103@vols.utk.edu,jysun@utk.edu
}
}


%Telephone: (800) 555--1212, Fax: (888) 555--1212}


\maketitle

\pagestyle{plain}

\begin{abstract}
%\boldmath
Leveraging users' behavioral data sampled by various sensors during the identification process, implicit authentication (IA) relieves users from explicit actions such as remembering and entering passwords. Various IA schemes have been proposed based on different behavioral and contextual features such as gait, touch, and GPS. However, existing IA schemes suffer from false positives, i.e., falsely accepting an adversary, and false negatives, i.e., falsely rejecting the legitimate user due to users' behavior change and noise. To deal with this problem, we propose BubbleMap (BMap), a framework that can be seamlessly incorporated into any existing IA system to balance between security (reducing false positives) and usability (reducing false negatives) as well as reducing the equal error rate (EER). To evaluate the proposed framework, we implemented BMap on five state-of-the-art IA systems. We also conducted an experiment in a real-world environment from 2016 to 2020. Most of the experimental results show that BMap can greatly enhance the IA schemes' performances in terms of the EER, security, and usability, with a small amount of penalty on energy consumption.

\end{abstract}


% \fi
%\IEEEpeerreviewmaketitle

%\vspace{-0.77cm}

\section{Introduction}
Implicit authentication (IA) is a promising method of authentication since it generally does not require any form of explicit user actions as in password, biometric, or token-based explicit authentication methods \cite{yang2016personaia,p9,yang2020dynamic}. Essentially, IA is achieved by matching users' historical behavior with their real-time behavior. Users' behavior is captured by various sensors embedded in smart devices, where the unique behavior pattern is extracted for each user. Historical behavior for a user can be derived by sensors whose data uniquely characterizes the user and distinguishes them from other users. It is updated after new data is available. In contrast, real-time behavior can be derived by the same sensors at the time of authentication. For energy saving purposes, IA schemes usually run in the background and stream data at an optimized frequency to ensure that data is sufficiently collected.

As with any other practical security system, IA systems need to strike a good balance between security and usability. On the one hand, we need the system to cope with the legitimate user's behavior deviation and noise \cite{yang2016personaia}, e.g., a change of routine, and not falsely rejecting the user (\emph{usability}). On the other hand, the system needs to differentiate between a legitimate user's behavior deviation and illegitimate users' behaviors to prevent falsely allowing the adversaries to access the system (\emph{security}). Compared to explicit authentication, IA is more susceptible to false negatives (falsely rejecting a legitimate user) and false positives (falsely accepting illegitimate users), due to the complexity of human behaviors and limitations of the machine learning algorithms used for extracting the user behavior model. It decreases the authentication accuracy IA systems are able to achieve and hinders the systems' wide deployment. In addition, since smart devices are a popular platform IA systems run on, energy consumption is an important consideration factor.

The majority of the existing IA schemes \cite{p7,bo2013silentsense,feng2014tips,shi2011senguard,shahzad2013secure,castelluccia2017towards,frank2013touchalytics,yang2020dynamic} tend to use a specific feature such as touch, typing, and location to uniquely identify users. However, due to the diversity of human behaviors, multiple features are needed to best identify each user in practice. For example, some people who take the train at a specific time during the weekdays will have a high user identification accuracy using location-based features, e.g., GPS. Since this situation may not be suitable for all users, e.g., the one who has unordered location patterns, a comprehensive system that adopts multiple features is necessary, especially in real usage.

%IA systems need to strike a good balance between security and usability. However, it is highly challenging to achieve such balance due to the dynamically changing behaviors of users. On the one hand, we need the system to cope with a user's behavior deviation \cite{yang2016personaia}, e.g., a change of routine, and not falsely reject the user (usability). On the other hand, the system needs to differentiate between a legitimate user's changed behavior and other users' behaviors to prevent falsely accepting adversaries (security). Balancing between such false rejects and false accepts improves the authentication accuracy and is the main focus of this paper.

In this paper, we propose BubbleMap (BMap), a framework to dynamically adjust and map users' privileges for accessing smart devices. This framework is independent of the machine learning algorithm and the features used in the IA scheme, and can be adopted by any existing IA scheme with various features \cite{p62,p7,bo2013silentsense,feng2014tips,shi2011senguard,gurary2016implicit,shahzad2013secure,castelluccia2017towards,p23, p7,frank2013touchalytics}, including the ones that run on wearable devices \cite{ekiz2019your,vhaduri2019multi}. It serves as a plug-in to reduce the equal error rate (EER) as well as to balance between security and usability. We introduce intermediate privilege levels to the two-level (full access or no access) systems used by the existing IA schemes in the \emph{Initial Mapping} step of BMap. An ideal IA scheme should always map the legitimate user to the top level (full access), and illegitimate users to the bottom level (no access). This paper extends our preliminary work \cite{yang2020dynamic} in terms of both theoretical design and real-world implementation. Specifically, we address the problems of illegitimate bubble expansion\footnotemark \footnotetext{The illegitimate bubble is defined in Section 2.2. We discuss \emph{Bubble Expansion} in Section 2.4. We discuss the other bubbles' movement in Section 2.3.} (Sections 2.2 and 2.4) and the other bubbles' movement (Section 2.3). In addition, we implement the proposed methods and conduct real-world experiments to evaluate the proposed system.

Nevertheless, when the legitimate user's behavior changes, it becomes harder to distinguish it from illegitimate users' behaviors. The intermediate privilege levels are incorporated in BMap to buffer the impact of behavior deviation so that the legitimate user still has access to some of the content (e.g., apps with lower security requirements), and therefore usability of the system is improved. It enhances the security of the system as well in that an adversary will not likely be mapped to the top level immediately due to the buffering. However, the ultimate goal of BMap is to help the IA system quickly reach a definitive conclusion by mapping the legitimate user to the top level and illegitimate users to the bottom level. This is achieved by the remaining steps of BMap, \emph{Privilege Movement} and \emph{Bubble Expansion} in which questions including where to move the current user's privilege, and how fast and how much to move it are addressed to reflect users' dynamically changing behavior. As shown in our experiments, the time users spend in the intermediate levels is only 0.4\% of their total usage time, and the equal error rate of the tested IA systems enhanced with BMap is universally decreased. Main contributions of this paper include:

$\bullet$ We design BMap and apply it to various state-of-the-art IA systems to boost their performance in terms of the equal error rate (EER), security, and usability. It is a plug-and-play that bridges the gap between IA research and the deployment of IA in practical systems.

$\bullet$ We implement BMap using Android smartphones and multiple servers. It contains various IA modules, such as data collection, privilege control, and user authentication.

$\bullet$ We evaluate the performances of the existing IA systems enhanced with BMap using large-scale comprehensive simulations and real-world experiments conducted over four years. We give quantitative results on the decrease of the equal error rate and analysis on improved security and usability. In addition, the energy consumption incurred by BMap is shown to be small.

\section{The Proposed BMap Framework}
We first provide an overview of the BMap framework and then elaborate on the detailed design.

\subsection{System Overview \label{sectionSystemOverview}}

\begin{figure}[htb]
%\vspace*{-1.2cm}
\begin{minipage}{1.0\linewidth}
\centering
  \begin{subfigure}[a]{\linewidth}
  \hspace*{-0.2cm}
    \centering
    \includegraphics[width=3.5in,height=0.7in]{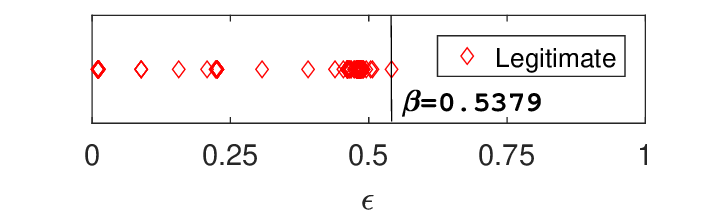}
    \caption{}\label{fig:evaPart1a}
  \end{subfigure}%

  \begin{subfigure}[b]{\linewidth}
  \hspace*{-0.2cm}
    \centering
    \includegraphics[width=3.5in,height=0.7in]{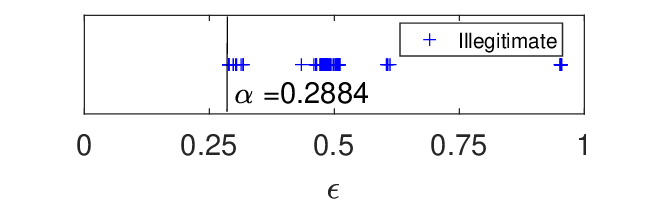}
    \caption{}\label{fig:evaPart1b}
  \end{subfigure}%
\end{minipage}%
  \caption{Behavior scores of (a) a legitimate user, (b) an illegitimate user.}
  \label{fig:evaPart1}
\end{figure}

Generally speaking, existing IA schemes authenticate users by deriving a behavior score $\epsilon$ using data gathered in a period of time, called time window (or authentication cycle) which is a design-specific parameter. Note that we also use the time window to derive a behavior score, where a time window contains multiple samples. The number of samples in each time window is dynamically adjusted based on the Wind Vane algorithm \cite{Yang2016}. Another purpose of applying the Wind Vane algorithm is to enable the template update ability \cite{pisani2019adaptive} of the original scheme and provide a fair comparison between original schemes and BMap boosted schemes. The average score $\epsilon$ derived by these samples and associated vectors is then compared with a threshold, e.g., $0.5$, and if the threshold is exceeded, the system concludes that the current user is illegitimate and locks the device. When legitimate and illegitimate users have vastly different behaviors, existing IA schemes can achieve low equal error rates \footnotemark. \footnotetext{The equal error rate (EER) represents the point where the false acceptance rate ($\frac{FA}{FA+TR}$) and false rejection rate ($\frac{FR}{FR+TA}$) are equal, where the true accept (TA) denotes a legitimate user's data sample has been correctly identified, otherwise denoted by false accept (FA), and the false reject (FR) denotes a legitimate user's data sample has been incorrectly identified to be illegitimate users' data sample, otherwise denoted by true accept (TR) \cite{tronci2009dynamic}. A good scheme should keep this value as small as possible.} However, based on our preliminary experiments using the Friends and Family dataset \cite{p21,aharony2011social}, more than 70\% of users' behavior data overlap and cannot be separated by simply setting a threshold. As a simple example, we randomly selected two participants from the dataset, one as the legitimate user and the other as the illegitimate user, and converted the system's output to probabilistic behavior scores \footnotemark. \footnotetext{In this test, the setting and features of the dataset are the same as Section \ref{pe}.} The time window is set to 15 seconds. The machine learning model is SVM with RBF kernel. We adopted ten-fold cross-validation during the test.  As shown in Fig. \ref{fig:evaPart1} (a) and (b), the legitimate and illegitimate users both have a large proportion of behavior scores located around the threshold of 0.5, which makes them inseparable. The behavior overlapping problem can be exacerbated by mimicry attacks where the adversary imitates the legitimate user's behaviors \cite{khan2016targeted}. BMap attempts to reduce the EER even in the presence of this problem by using the proposed \emph{Initial Mapping}, \emph{Privilege Movement}  and \emph{Bubble Expansion}, which will be discussed in what follows. In addition, we calculated the average percentage of overlapped behavior scores for every 130 participants in the dataset. The calculation shows that the average percentage of overlapped behavior scores among each legitimate and illegitimate user pair is 72\%. Furthermore, we conducted a statistical significance test using a one-sample t-test given a significance level of 0.05 with a null hypothesis: ``$H_0$: the overlapping percentage mean of behavior scores for each legitimate and illegitimate user pair equals 72\%". The result does not reject the null hypothesis with a $p-value=0.7634$. Hence, the 72\% overlapping ratio is statistically significant.

\begin{figure}[!h]
%\captionsetup[figure]{skip=-10cm}
%\setlength{\belowcaptionskip}{-10pt}
%\captionsetup{aboveskip=-3pt,belowskip=-99pt}
%\captionsetup{skip=-10pt}
\centering
\hspace*{-0.4cm}
\includegraphics[width=3.6in,height=2.5in]{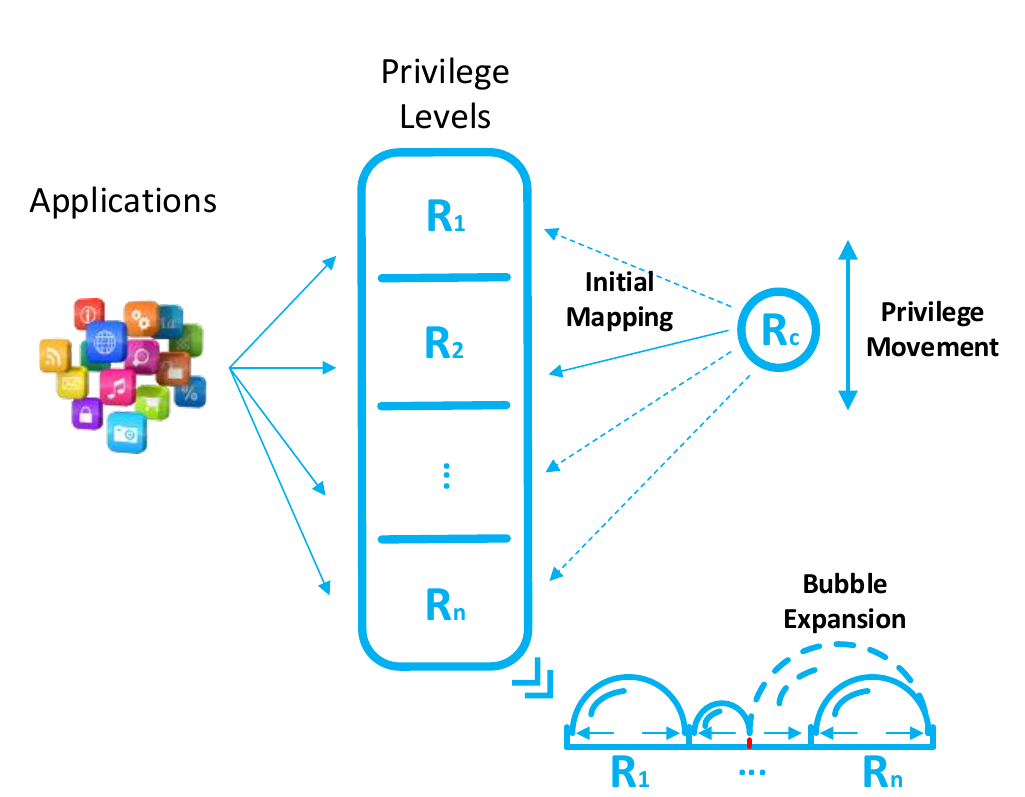}
% where an .eps filename suffix will be assumed under latex,
% and a .pdf suffix will be assumed for pdflatex; or what has been d8eclared
% via \DeclareGraphicsExtensions.
%\vspace{-0.3cm}
\caption{The BMap Framework.}
%\vspace{-0.3cm}
\label{fig:Sys}
\end{figure}

On a high level, in \emph{Initial Mapping}, three bubbles, legitimate bubble, slack bubble, and illegitimate bubble are created by defining their boundary values $\alpha$ and $\beta$ based on the legitimate user's historical behavior data. Each bubble contains different privilege levels representing access rights to apps of different security levels. \emph{Privilege Movement} then determines where to move the user's privilege level based on his current behavior. \emph{Bubble Expansion} is used to fine-tune the bubble sizes defined in \emph{Initial Mapping} and how far the privilege level should be moved in \emph{Privilege Movement}. It reduces the impact of noisy data and behavior deviation and therefore reduces the EER. Specifically, the first step in BMap is to map multiple privilege levels to the three bubbles using \emph{Initial Mapping}. We add a few intermediate privilege levels to the two-level (full access or no access) systems used by the existing IA schemes. Apps are categorized based on their security requirements and mapped to privilege levels. For instance, in a system with $n$ $(n \geq 3)$ privilege levels $R_1$ through $R_n$, apps can be mapped to the levels as shown in Fig. \ref{fig:Sys}. Apps with the highest security requirements such as banking, e-commerce, health and fitness, credit score, and password manager are mapped to the highest privilege level $R_1$. Apps with lower security requirements such as social media, texting, games, and utility apps are mapped to lower levels such as $R_2$, $R_3$, $R_{n-1}$. $R_n$ is the lowest privilege level which corresponds to locking the device with no access. The legitimate bubble, slack bubble, and illegitimate bubble contain the top level $R_1$, the intermediate levels $R_2$ to $R_{n-1}$, and the bottom level $R_n$, respectively. Note that defining the privilege levels, the security requirements for the apps, and their correspondence is system and user-dependent. It is relevant but not the focus of this paper. Interested readers are referred to \cite{sandhu1993lattice, ferraiolo2001proposed, yi2001security,hayashi2012goldilocks} for more details. Generally, a three-level system is required, and the number of privilege levels is proportional to the number of categories for different apps \cite{hayashi2012goldilocks}, e.g., utilities, entertainment, and news.  After obtaining the privilege levels, the system needs to map the user to a specific level $R_c$ based on the user's current behavior at the time of authentication. The level $R_c$ is called the user's current level, as shown in Fig. \ref{fig:Sys}. This is performed in the second step of BMap, \emph{Privilege Movement}. Once in this level, the user has access to all the apps corresponding to $R_c$ and the levels below $R_c$, but not the levels above $R_c$. Moreover, overlapping behaviors are effectively separated in this step. Finally, \emph{Bubble Expansion} is used to dynamically adjust the privilege boundaries as more behavior data becomes available and to filter out behavior and sensor noises. The bubble's status is similar to the real world one, which can be represented by using physical laws.

\textbf{Adversaries in BMap} BMap defends against password guessing attacks and behavior mimicry attacks (adversaries imitating legitimate users' behavior). In password guessing attacks, we assume that it takes the adversary a few tries (exceeding the limit) to guess and input the password correctly. Likewise, it requires some time for the adversary to fully mimic the legitimate user's behavior\cite{khan2016targeted}. We do not consider an adversary who can enter the password correctly within the trial limit because this is a general problem common to all existing authentication systems using passwords. Without loss of generality, we assume the use of a backup authentication mechanism to count for when the IA fails, similar to how a passcode is required (with limited tries) to unlock an iPhone X when Face ID fails. A password-guessing adversary may eventually correctly guess and input the passcode. But due to \emph{Bubble Expansion}, every wrong password input accelerates the expansion of the illegitimate bubble causing the system to quickly map the adversary to the bottom level, thus reducing false positives. Since mimicry attacks also require launch time \cite{khan2016targeted}, BMap defends against them in a similar manner. BMap also reduces false negatives by quickly mapping the legitimate user's privilege back to the top level once a correct passcode is entered within reasonable tries. We use the terms illegitimate user and adversary interchangeably.

The passcode used to unlock the device is not to be confused with the PIN number, which the user can enter to re-initialize the system and retrain the machine learning model based on new behavior data\cite{p67}. This can be a useful feature when the system keeps failing the legitimate user's authentication and needs calibration.

%The reason for such design is that since implicit authentication systems are still in their infancy, understanding their performance limitations is the most important first step before we can mature their design. The second factor serves as a feedback mechanism in BMap to help separate behavior deviation from illegitimate behaviors, and fundamentally improve the system's false accept and false reject rates. Despite that we use password input as the second factor in this paper, any authentication scheme other than behavior-based implicit authentication can be used.

In this paper, we use a subset of all available features in the Friends and Family dataset, i.e., GPS, accelerometer, touch, SMS, app installation, battery usage, call logs, app usage, blue-tooth devices log, and Wi-Fi access points to evaluate the existing IA schemes. For example, we use the accelerometer data for Gait \cite{ravi2005activity} and the touch and accelerometer data for SilentSense \cite{bo2013silentsense}. In order to give a fair comparison, we applied BMap to five state-of-the-art IA schemes, Shi \cite{p9}, Multi-Sensor \cite{p62}, Gait \cite{ravi2005activity}, SilentSense \cite{bo2013silentsense}, and Touchalytics \cite{frank2013touchalytics}, by strictly following the feature selection and parameter tuning process in these schemes.

%The majority of current implicit authentication research tends to gather their own data from a small number of volunteers \cite{khan2016targeted,p65,bo2013silentsense,p62}, rendering it difficult to repeat their tests. We use the public Friends and Family Dataset that contains comprehensive user behavior data for the presentation and evaluation of our BMap. Specifically, the dataset contains 130 participants' 8GB data collected in a 5-month period. Data consists of 9 main features: GPS, accelerometer, SMS, app installation, battery usage, call logs, app usage, blue-tooth devices log, and Wi-Fi access points. Some of them have many sub-features, e.g., battery usage includes battery level, plug status, health and brand information. In our tests, we randomly select one user as the legitimate user and use other users' data as illegitimate behavior data.

\subsection{Defining Bubbles \label{ilm}}

We mainly discuss applying BMap to SVM-based IA schemes \cite{p62,ravi2005activity,bo2013silentsense,frank2013touchalytics}. For the other IA schemes \cite{p14,yang2016personaia, shi2011senguard}, since their output is already a probabilistic behavior score, BMap can be directly applied.

\vspace{0.12cm}

DEFINITION 1. \emph{Let behavior score $\epsilon \in$ [0, 1] denote the probabilistic output of an SVM approximated by a two-parameter sigmoid function $\frac{1}{1+exp(Af_i+B)}$. In a specific training set \footnotemark, we further divide the interval [0, 1] into $n$ sub-intervals, called bubbles, denoted by $D_n\subset$ [0, 1].}\footnotetext{A training set is a dataset that contains various users' historical behavioral data.}

%\vspace{0.12cm}

\begin{figure}[htb]
\vspace*{-0.5cm}
\centering
  \begin{subfigure}[b]{1\linewidth}
  %\hspace*{-0.9cm}
    \centering
    \includegraphics[width=3.5in,height=1.9in]{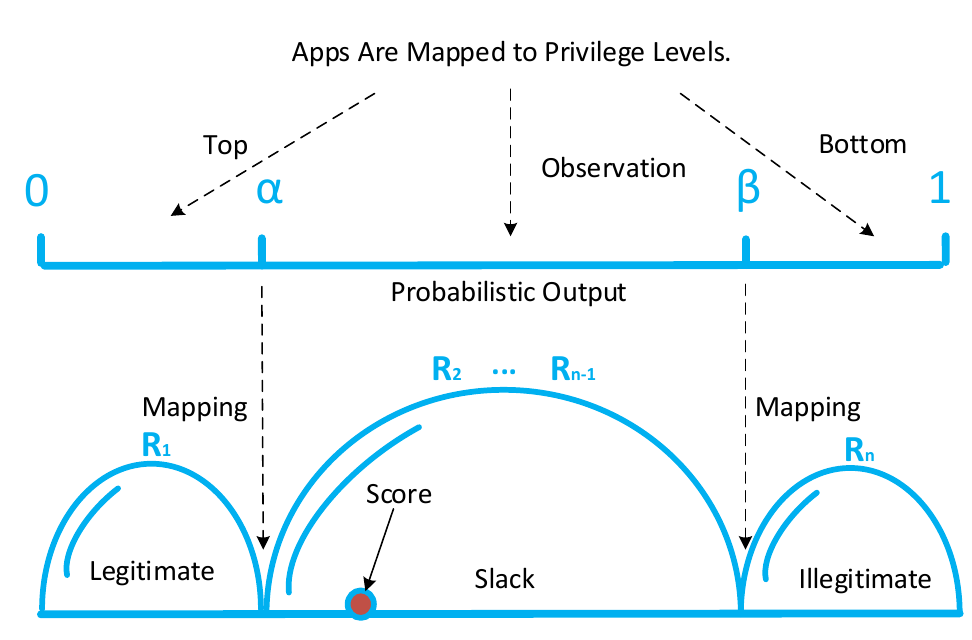}
    %\caption{}\label{fig:arca}
  \end{subfigure}%
  \caption{Initial mapping.}
  \label{fig:InitialMapping}
\end{figure}

The \emph{Initial Mapping}  mechanism is illustrated in Fig. \ref{fig:InitialMapping}. The system first initializes the value of parameters $\alpha$ and $\beta$ by fitting the sigmoid function to the SVM output trained by data sampled from legitimate and illegitimate users \cite{platt1999probabilistic}. Note that the distance between $\alpha$ and $\beta$ can be very small, e.g., 0.01, but they never collide.

%\vspace*{1.20cm}

DEFINITION 2. \emph{In various bubbles, the legitimate bubble is the largest sub-interval that contains only true accept (TA) behavior scores. The illegitimate bubble is the largest sub-interval that contains only true reject (TR) behavior scores. The slack bubble is the sub-interval in between the legitimate bubble and the illegitimate bubble.}

The legitimate, slack, and illegitimate bubbles are blown based on these two parameters, in which only the legitimate bubble can explode. Assuming the system has $n$ privilege levels, in each authentication cycle as new data is collected, the SVM takes the data as input and outputs a new behavior score indicating the system's authentication decision. If the new score falls in the legitimate bubble, the system will move the user's current privilege level $R_c$ to $R_1$ (if $R_c\neq R_1$), which grants the user full access. If the new score falls in the illegitimate bubble, the system will lock the device. If the new score falls in the slack bubble, the system will map $R_c$ to one of the observation levels $R_2$, $R_3$, ..., $R_{n-1}$, where the user has only limited access.

%The identification results are shown in Fig. \ref{fig:eva}, in which each point represents a behavior score during the predefined time window. As shown in Fig. \ref{fig:eva}, both legitimate and illegitimate users have a large proportion of behavior scores located around $p=0.5$. If we set the threshold to be 0.25, it increases the false reject rate. If we set the threshold to be 0.55, it increases the false accept rate. The only reasonable choice is to select a point in between 0.4 to 0.6, but we can not achieve a very high accuracy during the authentication phase. Furthermore, if we want to increase the security we should lower the threshold that precludes any samples from illegitimate user, but it potentially increase the chance of locking the device. On the other hand, if we want to increase the usability we should extend threshold, but it potentially increases the by-pass rate.

As shown in Fig. \ref{fig:evaPart1} (a) and (b), the legitimate and illegitimate bubbles are [0, $\alpha$] and [$\beta$, 1], respectively. The slack bubble is located in [$\alpha$, $\beta$], which contains ambiguous behavior scores that could come from either the legitimate user or illegitimate users and need separation. In a given dataset, we can easily find $\alpha$ and $\beta$ by searching for the largest and smallest behavior score $\epsilon$ derived from the legitimate user's and illegitimate users' training data, respectively. In \emph{Initial Mapping}, we first assume that $\alpha$ and $\beta$ are fixed and focus on the mapping of the current privilege level $R_c$ to one of the observation levels in the slack bubble. We then release this assumption in Section \ref{da} when we complete our discussion with the possible movement of the bubble boundaries. Compared to the existing implicit authentication schemes, \emph{Initial Mapping} in BMap focuses on both security and usability. Since the system only grants full access to the user who is most likely to be legitimate, security is enhanced. When the likelihood declines, instead of completely locking the user out, the system maps the user to an observation level that grants lower access rights. It enhances usability if the user is legitimate while limiting the security breach if the user is illegitimate. Nevertheless, \emph{Initial Mapping}  only handles failed authentications in a more gradual way by adding the slack bubble and observation levels. It does not fundamentally ameliorate the false reject (FR) and false accept (FA) performance, which will be the focus of \emph{Privilege Movement} and \emph{Bubble Expansion}.

In \emph{Initial Mapping}, the current privilege $R_c$ is mapped to one of the defined privilege levels $[R_1, R_2, ..., R_n]$ when a new behavior score becomes available at the time of authentication and remains in that level until more data comes in. Such a mapping mechanism does not fundamentally improve the FR and FA performance since the system still needs a way to confirm the user's legitimacy once her behavior score is mapped to the uncertain observation level. Recall that the system's goal is to eventually grant the user full access if she is legitimate and lock her out otherwise. The slack bubble is just a buffer for a smoother transition. We introduce \emph{Privilege Movement} in the mapping of $R_c$, where $R_c$ is moved up (towards $R_1$) or down (towards $R_n$) gradually out of the slack bubble. We assume that the implicit authentication scheme gives a low EER, i.e., the legitimate and illegitimate users' behavior scores fall into their corresponding bubbles rather than the slack bubble when the scheme is newly trained.

\subsection{Movement Directions \label{dj}}
\begin{figure}[htb]
%\vspace*{-0.5cm}
\centering
  \begin{subfigure}[b]{1\linewidth}
  %\hspace*{-0.5cm}
  \vspace*{-0.20cm}
    \centering
    \includegraphics[width=3.5in,height=2.5in]{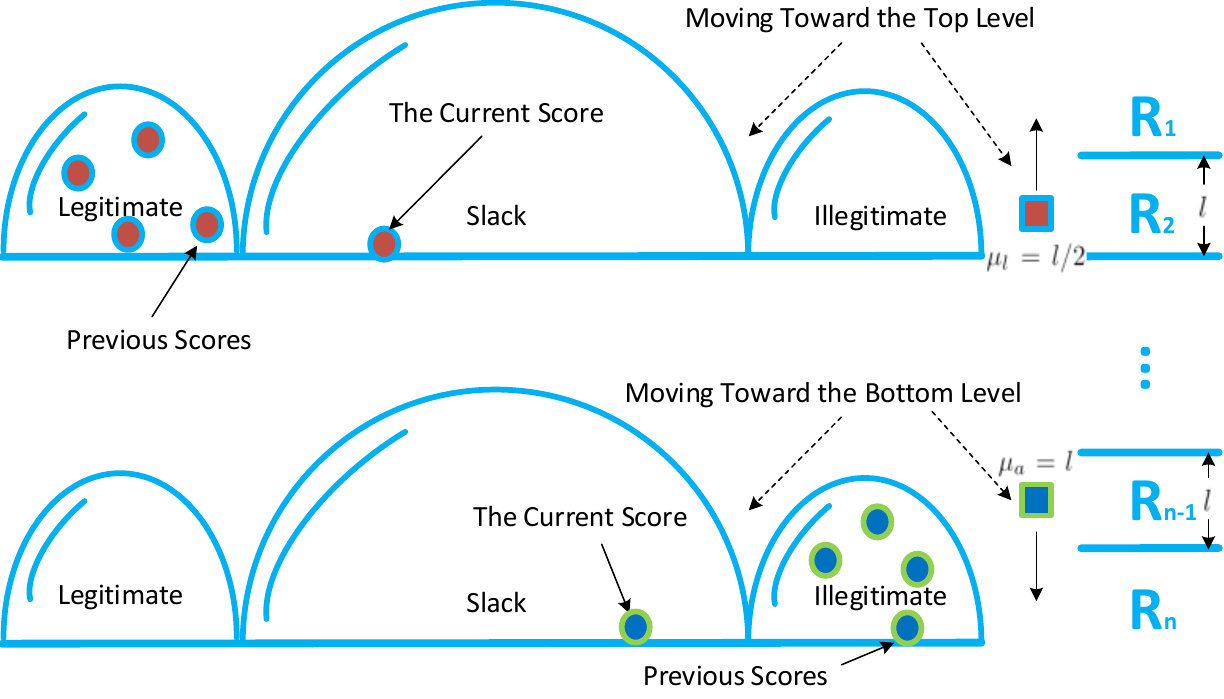}
    %\caption{}\label{fig:arcc}
  \end{subfigure}%
  \caption{Privilege movement.}
  \label{fig:PrivilegeMovement}
\end{figure}
%\vspace*{1.20cm}

%Authentication accuracy will gradually decline as more behavior data becomes available from either the legitimate user or illegitimate users after training. Retraining of the implicit authentication scheme may be needed which is covered in detail in \cite{p67}.

\begin{figure*}[htb]
%\vspace*{-1.2cm}

  \begin{minipage}{.33\textwidth}
  \hspace*{-0.2cm}
  \begin{subfigure}[c]{\linewidth}
  %\hspace*{-0.2cm}
    \centering
    \includegraphics[width=1.099\textwidth,height=2.4in]{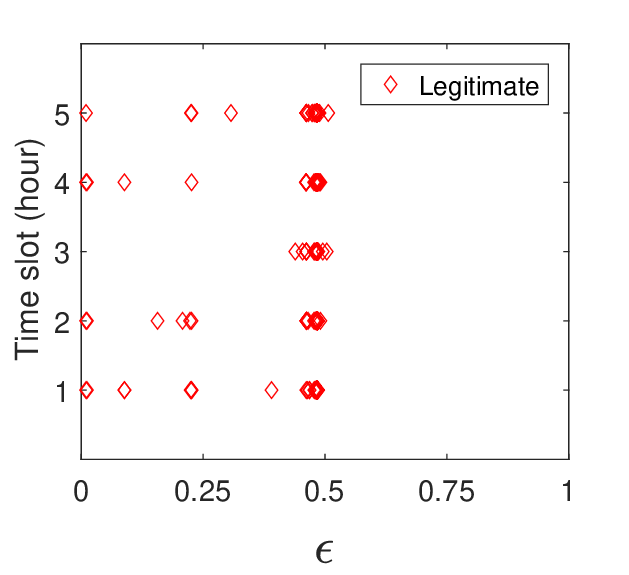}
    \caption{}\label{fig:evac}
  \end{subfigure}%
\end{minipage}%
\begin{minipage}{.33\textwidth}
  \hspace*{-0.2cm}
  \begin{subfigure}[d]{\linewidth}
  %\hspace*{-0.2cm}
    \centering
    \includegraphics[width=1.099\textwidth,height=2.4in]{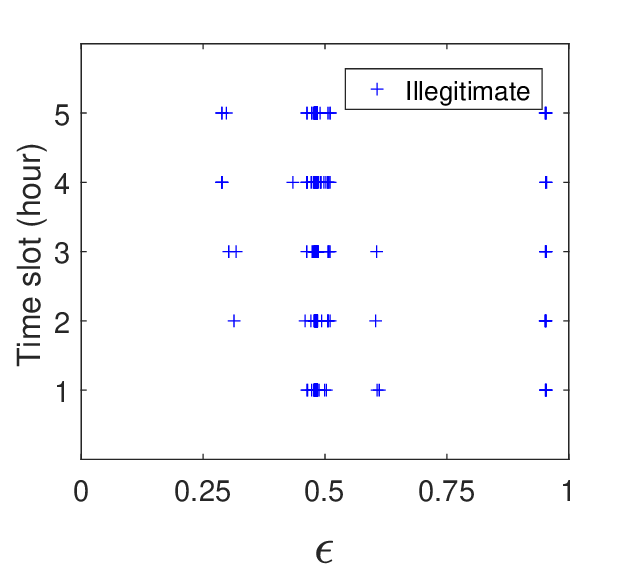}
    \caption{}\label{fig:evad}
  \end{subfigure}%
\end{minipage}%
\begin{minipage}{.33\textwidth}
  \hspace*{-0.2cm}
  \begin{subfigure}[e]{\linewidth}
  %\hspace*{-0.2cm}
    \centering
    \includegraphics[width=1.099\textwidth,height=2.4in]{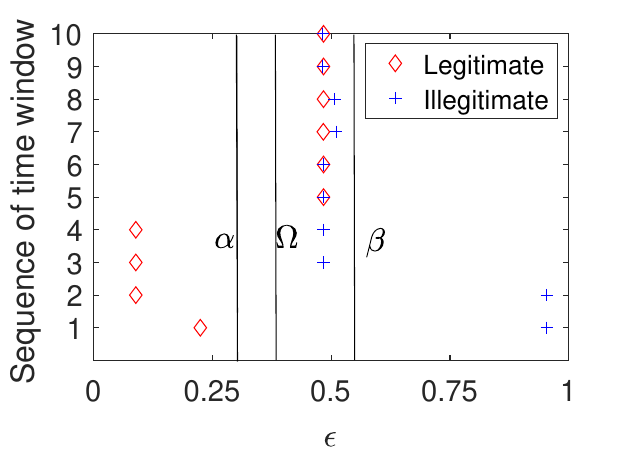}
    \caption{}\label{fig:evae}
  \end{subfigure}%
\end{minipage}%
  \caption{Behavior scores of (a) the legitimate user in a 5-hour period, (b) illegitimate users in a 5-hour period, and (c) both users in 10 time windows.}
  \label{fig:eva}
\end{figure*}

We summarize \emph{Privilege Movement} mechanism in Fig. \ref{fig:PrivilegeMovement}. The system keeps tracking the user's behaviors and once it observes a behavior score that falls into the slack bubble, it searches through the previous scores to find a more definitive answer. If there were scores in the legitimate bubble, the system leans towards regarding the user as legitimate and moves $R_c$ upward with distance $-\mu_l$ at the end of the current authentication cycle. This process is repeated until $R_c$ reaches $R_1$, if the score keeps falling into the legitimate bubble. Similarly, if there were scores in the illegitimate bubble, the system leans towards regarding the user as illegitimate and moves $R_c$ downward with distance $+\mu_a$ at the end of the current authentication cycle. This process is repeated until $R_c$ reaches $R_n$, if the score keeps falling into the illegitimate bubble. If $R_c$ falls in between privilege levels, the user is assumed to access privilege of the lower level. The movement distances $-\mu_l$ and $+\mu_a$ are design parameters that can be constants or variables. For the discussion in this subsection, we let $\mu_l=l/2$ and $\mu_a=l$ where $l$ is the fixed distance between two privilege levels. The system is thus less tolerable and more restrictive when there is evidence that the current user is illegitimate. It is also more conservative in giving the user higher access privilege when the user's legitimacy was confirmed in the past but is currently in doubt. Such design is to enhance security while not sacrificing usability. Moreover, the FR and FA performance is improved since the system always tries to move $R_c$ out of the slack bubble based on evidence. \emph{Privilege Movement} mechanism has $O(1)$ time complexity, which renders the system's latency the same as the implicit authentication schemes without BMap. In the next subsection, we discuss making $\mu_l$ and $\mu_a$ variables to reduce the EER.

As an example, the behavior score distribution for the legitimate user and illegitimate user is shown in Fig. \ref{fig:eva} (a) and (b), respectively, using the aforementioned simulation setting and data from two participants (Fig. \ref{fig:evaPart1} in Section \ref{sectionSystemOverview}). The scores are grouped into five one-hour time slots, where each time slot contains multiple time windows; and each time window contains multiple samples.
In each time slot, there are behavior scores belonging to the legitimate/illegitimate bubble that co-occur with scores belonging to the slack bubble. The scores that belong to the legitimate/illegitimate bubble are used as evidence and guidance to move the scores in the slack bubble. When behavior deviation happens, \emph{Initial Mapping}  may map the legitimate user to the observation level and still cause false rejects which are corrected with \emph{Privilege Movement}. The same is true for false accepts. In addition, we randomly selected a time slot from Fig. \ref{fig:eva} (a) and (b), and magnified it in Fig. \ref{fig:eva} (c), where the threshold $\Omega$ is predefined to best separate the two users. For the ease of presentation, we assume that there is only one observation level and three privilege levels in total. In the first time window, the legitimate user's behavior score falls in the legitimate bubble (shown in the figure), but her $R_c$ has not reached $R_1$ (not shown in the figure). The system, therefore, moves $R_c$ upward for $l/2$. In the second through fourth time windows, the score falls in the legitimate bubble again but $R_c$ has reached $R_1$. So $R_c$ remains in $R_1$. In the fifth through tenth time windows, $R_c$ falls in the slack bubble. Since the system observed four behavior scores in the legitimate bubble, $R_c$ remains in $R_1$. If the system observed scores in the illegitimate bubble instead, $R_c$ would have been moved towards $R_n$. The illegitimate user in Fig. \ref{fig:eva} (c) follows a similar Privilege Movement process. Using the dataset \cite{p21}, we were able to observe the co-occurrence of legitimate/illegitimate-bubble behavior scores and slack-bubble behavior scores for the same user in a reasonably short period of time (2-3 minutes), in all of the two-participant simulations we conducted. For ease of understanding, a simplified version of \emph{Privilege Movement} mechanism is shown in Alg. \ref{alg1}, where the movement direction and distance are guided by the most recent score that falls in the legitimate/illegitimate bubble. In practice, instead of using only one score, we utilize multiple scores to make a more accurate decision.

\begin{algorithm}[h]
\KwIn{A Behavior Score}
\KwOut{A Privilege Movement Decision, U}
\nl initialize S:=A Behavior Score \;
\nl initialize U:=0 \;\tcc{\footnotesize{Movement direction and distance.}}
\nl \If{(S$<=$alpha \& S$>=$0)}{\tcc{\footnotesize{Falling in the legitimate bubble.}}
\nl U=$-l/2$\;
\nl direction=-1\;\tcc{\footnotesize{Toward the legitimate bubble.}} 
\nl time\_gap=0\;\tcc{\footnotesize{time\_gap represents the number of time windows between now and the last time we saw the score falling in the legitimate/illegitimate bubble.}}
}
\nl \ElseIf{(S$>=beta$ \& S$<=$1)}{\tcc{\footnotesize{Falling in the illegitimate bubble.}}
\nl U=$l$\;
\nl direction=1\;\tcc{\footnotesize{Toward the illegitimate bubble.}}
\nl time\_gap=0\;
}
\nl \Else{\tcc{\footnotesize{Falling in the slack bubble.}}
\nl    \If{(time\_gap$<=$max\_gap)}{
\nl        \If{(direction$<$0)}{
\nl             U=-$l/2$\;
            }
\nl        \Else{
\nl             U=$l$\;
            }
         }
\nl     \Else{
\nl        lock\_the\_device()\;         
        }
\nl     time\_gap ++\;        
}

\nl  \Return {} U\;
\caption{{\bf Privilege Movement (Simplified)} \label{alg1}}
\end{algorithm}

The effectiveness of \emph{Privilege Movement} is highly dependent on the size of the legitimate and illegitimate bubbles. If $\alpha$ and $\beta$ are fixed, they may become less indicative as more behavior data from either the legitimate user or illegitimate users become available. This problem will be addressed in the \emph{Bubble Expansion} mechanism, where the size of the bubbles is dynamically adjusted to reflect the behavior change and reduce the EER.

\subsection{Bubble Expansion \label{da}}
\begin{figure}[!h]
%\captionsetup[figure]{skip=-10cm}
%\setlength{\belowcaptionskip}{-10pt}
%\captionsetup{aboveskip=-3pt,belowskip=-99pt}
%\captionsetup{skip=-10pt}
\centering
%\hspace*{-0.9cm}
%\includegraphics[width=7in,height=2.2in]{FingerSysNew.pdf}
\includegraphics[width=3.5in,height=2.4in]{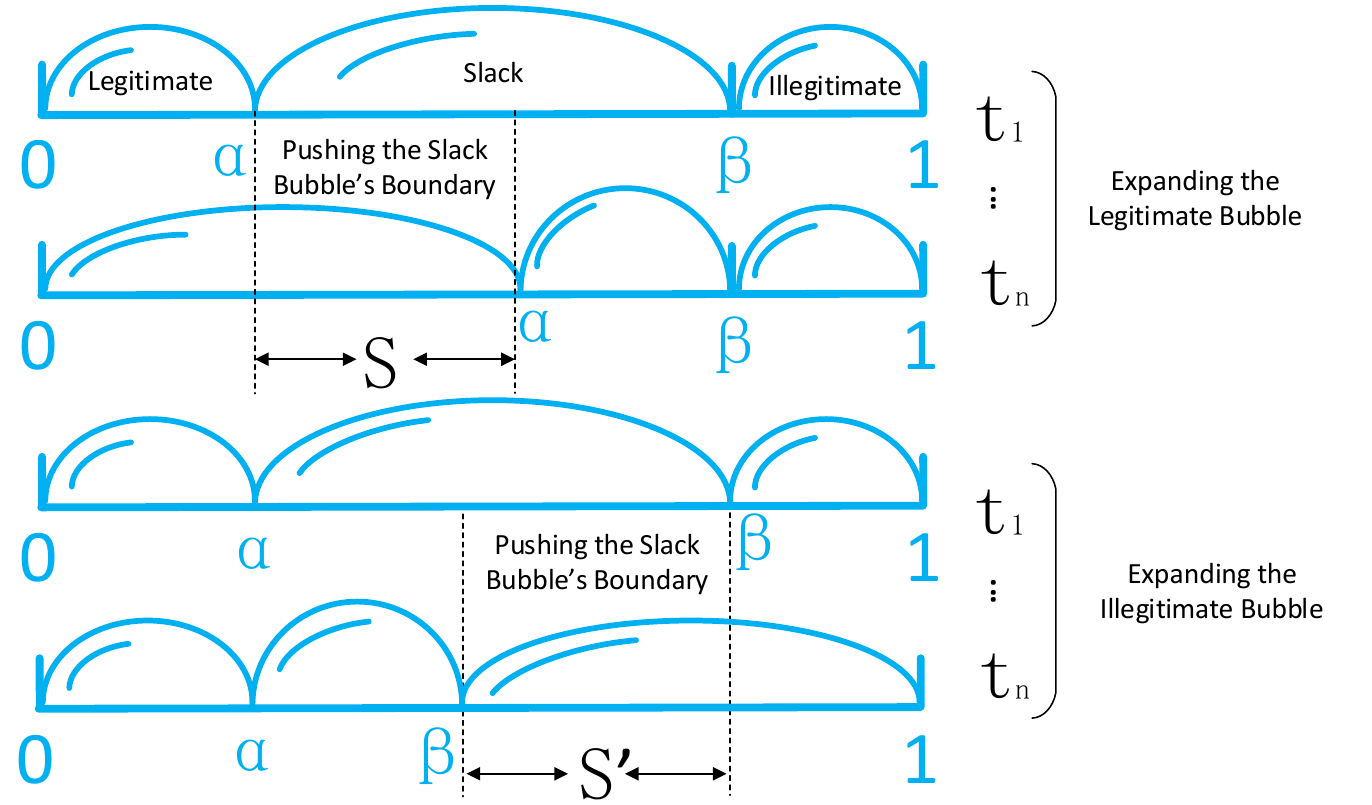}
% where an .eps filename suffix will be assumed under latex,
% and a .pdf suffix will be assumed for pdflatex; or what has been d8eclared
% via \DeclareGraphicsExtensions.
%\vspace{-0.3cm}
\caption{Bubble expansion.}
%\vspace{-0.3cm}
\label{fig:dj}
\end{figure}

We now introduce \emph{Bubble Expansion}, in which the bubble boundaries $\alpha$ and $\beta$ are updated. In practice, due to behavior deviation and sensor noise, the initial setting of $\alpha$ and $\beta$ may become inaccurate. If behavior scores from the legitimate user keep falling in the slack bubble, it may indicate that the legitimate bubble is too small and more ``air" is needed to reduce false rejects. Similarly, the illegitimate bubble may need to be expanded to reduce false accepts. The EER is reduced as a result. As shown in Fig. \ref{fig:dj}, the original legitimate and illegitimate bubbles are [0, $\alpha$] and [$\beta$, 1], respectively. The new bubbles become [0, $\alpha'$] and [$\beta'$, 1] after expansion. In addition, the system's latency is reduced since less \emph{Privilege Movement} is needed and the system can make decisions more quickly.

In a given dataset, it is straightforward to find out whether the behavior scores that keep falling in the slack bubble belong to the legitimate user. In reality however, it is difficult for the system to know in which case the second-factor authentication (password input for our discussion) is needed to provide feedback, as previously mentioned. We assume that the legitimate user will input the correct password and illegitimate users will input incorrect passwords at the beginning of usage. Although illegitimate users can guess passwords, after several unsuccessful tries the chance that illegitimate users are locked out is increased exponentially due to illegitimate bubble expansion. Similarly, an attacker can also mimic legitimate users' behavior, but it also requires time \cite{khan2016targeted}. Due to the \emph{Privilege Movement} mechanism, compared to original schemes, attackers will be blocked before they fully mimic legitimate users' behavior. For this reason, the true reject rate and the system's security are increased. Correspondingly, due to legitimate bubble expansion, legitimate users who input correct passwords at the beginning of usage will have a larger chance of being mapped to the top privilege level. For this reason, the true accept rate and system's usability are increased.

We model \emph{Bubble Expansion} by applying physical laws that describe the motion of bodies under the influence of a system of forces. Specifically, the expansion $S$ in time $t$ is defined as:
\begin{equation}
\label{fun3}
\begin{split}
%\begin{aligned}
S=\frac{1}{2}(\mathcal{a}-\hat{\mathcal{a}})t^2+v_0t,
\end{split}
\end{equation}
where $\mathcal{a}$ denotes the acceleration of the expansion, $t$ denotes the number of time windows or authentication cycles, $v_0$ denotes the initial velocity of the expansion (normally, we choose $v_0=0$), and $\hat{\mathcal{a}}$ is the resistance that slows down or stops the expansion. Every time the user inputs the correct password and the behavior score is outside of the legitimate bubble, more ``air" will be blown into the legitimate bubble, and expand it to contain the behavior score where the expansion is proportional to the distance between the behavior score and legitimate bubble ($\epsilon-\alpha$). The legitimate bubble can keep expanding and pushing the left boundary of the slack bubble to move toward the illegitimate bubble.

The acceleration of the expansion $\mathcal{a}$ is defined as:
\begin{equation}
\label{fun1}
\begin{split}
%\begin{aligned}
\mathcal{a}=\frac{R_d*\varepsilon}{W_1}+W_2+\delta,
\end{split}
\end{equation}
where $W_1=\frac{\sum_i n_l^{(i)}+n_a^{(i)}}{\sum_i N^{(i)}}$ is a balancing parameter that controls the expansion, $W_2$ is a constant representing the initial acceleration, $\sum_i n_l^{(i)}$ is the number of times the user inputs the correct password when her score is in the slack bubble, $\sum_i n_a^{(i)}$ is the number of times the user inputs a wrong password when her score is in the slack bubble, $\sum_i N^{(i)}$ is the total number of authentication cycles, $R_d$ is the distance between $R_c$ and $R_1$, $\varepsilon=\epsilon-\alpha$ is the distance between the behavior score and legitimate bubble, and $\delta$ is the mixture of behavior noise and sensor noise.

The expansion of the legitimate bubble may result in the inclusion of illegitimate users' behavior scores that originally fall in the slack bubble. To reduce such false accepts, we introduce the resistance $\hat{\mathcal{a}}$ that constrains the expansion:
\begin{equation}
\label{fun2}
\begin{split}
%\begin{aligned}
\hat{\mathcal{a}}=\mathcal{a}(\int_0^\alpha p(\varepsilon_a)d\varepsilon_a +\theta),
\end{split}
\end{equation}
where $\theta$ is a constant that prevents $\alpha$ from surpassing $\beta$, $\int_0^\alpha p(\varepsilon_a)d\varepsilon_a$ denotes the probability that the legitimate bubble contains behavior scores derived from illegitimate users in the training set, and $\varepsilon_a$ denotes the behavior score derived from illegitimate users' data in the training set. $\int_0^\alpha p(\varepsilon_a)d\varepsilon_a$ is estimated using kernel density estimator \cite{scott2008kernel,measurekernel,bishop2006pattern}.

Substituting (\ref{fun2}) into (\ref{fun3}) and assuming $t=1$, we have
\begin{equation}
\label{fun4}
\begin{split}
%\begin{aligned}
S=\frac{1}{2}\mathcal{a}(1-\int_0^\alpha p(\varepsilon_a)d\varepsilon_a-\theta)+v_0,
\end{split}
\end{equation}
where we let $V=1-\int_0^\alpha p(\varepsilon_a)d\varepsilon_a-\theta$, called fluid viscosity, control when the expansion stops.

Substituting \ref{fun1} into \ref{fun4}, we have
\begin{equation}
\label{fun5}
\begin{split}
%\begin{aligned}
S=\frac{1}{2}(\frac{R_d*\varepsilon}{W_1}+W_2)V+v_0+\Delta,
\end{split}
\end{equation}
where $\Delta=\frac{V*\delta}{2}$ is estimated and eliminated using a Kalman filter \cite{kalman}.

The aforementioned models (Eqn. 5) describe the expansion of the legitimate bubble. To describe the illegitimate bubble's expansion, we have 

\begin{equation}
\label{fun3'}
\begin{split}
%\begin{aligned}
S'=\frac{1}{2}(\mathcal{a'}-\mathcal{\hat{a}'})t^2+v_0t,
\end{split}
\end{equation}

where $S'$ denotes the total displacement, $\mathcal{a'}$ denotes the acceleration of the illegitimate bubble expansion, $t$ denotes the number of time windows or authentication cycles, $v_0$ denotes the initial velocity of the expansion (normally, we choose $v_0=0$), and $\mathcal{\hat{a}'}$ is the resistance that slows down or stops the expansion. Every time the user inputs the wrong password and the behavior score is outside of the illegitimate bubble, more ``air" will be blown into the illegitimate bubble, and expand it to contain the behavior score where the expansion is proportional to the distance between the behavior score and illegitimate bubble ($1 -\epsilon$). The illegitimate bubble can keep expanding and pushing the right boundary of the slack bubble to move toward the legitimate bubble. Then it will further push the legitimate bubble until it explodes ($\alpha=0$).  

The acceleration of the expansion $\mathcal{a'}$ is defined as: 

\begin{equation}
\label{fun1'}
\begin{split}
%\begin{aligned}
\mathcal{a'}=\frac{1-\varepsilon}{R_d*W_1}+W_2+\delta',
\end{split}
\end{equation},
where $1-\varepsilon$ denotes the distance between the behavior score and illegitimate bubble, $\delta'$ is the mixture of behavior noise and sensor noise.

The expansion of the illegitimate bubble may result in the inclusion of legitimate users' behavior scores that originally fall in the slack bubble. To reduce such false reject, we introduce the resistance $\mathcal{\hat{a}'}$ that constrains the expansion:

\begin{equation}
\label{fun2'}
\begin{split}
%\begin{aligned}
\mathcal{\hat{a}'}=\mathcal{a'}(\int_\beta^1 p(\varepsilon_l)d\varepsilon_l +\theta'),
\end{split}
\end{equation}
where $\theta'$ is a constant that prevents $\beta$ from surpassing $\alpha$, $\int_\beta^1 p(\varepsilon_l)d\varepsilon_l$ denotes the probability that the illegitimate bubble contains behavior scores derived from legitimate users in the training set, and $varepsilon_l$ denotes the behavior score derived from legitimate users' data in the training set. 

Substituting (8) into (6) and assuming t=1, we have

\begin{equation}
\label{fun4'}
\begin{split}
%\begin{aligned}
S'=\frac{1}{2}\mathcal{a'}(1-\int_\beta^1 p(\varepsilon_l)d\varepsilon_l-\theta')+v_0,
\end{split}
\end{equation}
where we let $V'=1-\int_\beta^1 p(\varepsilon_l)d\varepsilon_l-\theta'$, called fluid viscosity, control when the expansion stops.

Substituting (7) into (9), we have 

\begin{equation}
\label{fun5'}
\begin{split}
%\begin{aligned}
S'=\frac{1}{2}(\frac{1-\varepsilon}{R_d*W_1}+W_2)V'+v_0+\Delta'
\end{split}
\end{equation}

where $\Delta'=\frac{V'*\delta}{2}'$ is estimated and eliminated using a Kalman filter.

In each authentication cycle, if the user inputs the correct password, the predicted state estimate $x_{k|k-1}$ which controls the expansion of the legitimate bubble is defined as: $x_{k|k-1}=F_kx_{k-1|k-1}+B_ku_k$, where
$F_k=\begin{bmatrix}
    1 &t \\
    0 &1 \\
\end{bmatrix}
$, $B_k=\begin{bmatrix}
    \frac{t^2}{2}\\
    t \\
\end{bmatrix}$ and $u_k=(\frac{R_d*\varepsilon_a}{W_1}+W_2)V$. The predicted estimate covariance $P_{k|k-1}$ is defined as: $P_{k|k-1}=F_kP_{k-1|k-1}F_k^T+Q_k$, where the process noise covariance is $Q_k=\begin{bmatrix}
    \frac{t^4}{4} &\frac{t^3}{2} \\
    \frac{t^3}{2} &t^2 \\
\end{bmatrix}*\sigma_a^2$ with $\sigma_a$ being the magnitude of the process noise (behavior noise).  The innovation covariance is $S_k=H_kP_{k|k-1}H_k^T+R_k$, where $H_k=\begin{bmatrix}
    1 \\
    0 \\
\end{bmatrix}$ and $R_k$ is the covariance of the observation noise (sensor noise). Kalman gain is calculated as: $K_k=P_{k|k-1}H_k^TS_k^{-1}$. Since a Kalman filter is loop carried, we update the state estimate and associated covariance at the end of each authentication cycle as: $x_{k|k}=x_{k|k-1}+K_k(z_k-H_kx_{k|k-1})$, and $P_{k|k}=(I-K_kH_k)P_{k|k-1}$. We calculate the expansion as $P_{k|k}H_k$ and need to rescale it before applying it to real systems.

If the user inputs a wrong password, we let $u_k=\frac{\varepsilon_l}{R_d*W_1}+W_2$. Furthermore, to defend password guessing, if users continuously input wrong password $m$ times, the legitimate bubble will explode until users re-blow it by passing the hidden factor authentication discussed in Section \ref{sectionSystemOverview}. Instead, the illegitimate bubble will keep expanding and finally cause the legitimate bubble to shrink. Note that the bubble boundaries $\alpha$ and $\beta$ never collide, since the slack bubble could become very small, e.g., with length of 0.01, but it never explodes.

In addition to causing false accepts, the expansion of the legitimate bubble also affects \emph{Privilege Movement}, or more specifically, the distance of the movement $-\mu_l$ and $+\mu_a$. Now that the bubble boundaries $\alpha$ and $\beta$ are dynamically adjustable, the distance of the movement needs to be adjusted accordingly. Leveraged kernel density estimator \cite{scott2008kernel,measurekernel,bishop2006pattern}, we let $-\mu_l=-\mu_l\frac{\int_0^\alpha p(\varepsilon_l)d\varepsilon_l}{\int_0^\alpha p(\varepsilon_a)d\varepsilon_a}$ and $+\mu_a=+\mu_a\frac{\int_\beta^1 p(\varepsilon_a)d\varepsilon_a}{\int_\beta^1 p(\varepsilon_l)d\varepsilon_l}$, where $\varepsilon_l$ and $\varepsilon_a$ denote the behavior scores derived from the legitimate user's and illegitimate users' data in the training set, respectively; $\int_0^\alpha p(\varepsilon_l)d\varepsilon_l$ and $\int_0^\alpha p(\varepsilon_a)d\varepsilon_a$ denote the probabilities that the legitimate bubble contains behavior scores derived from the legitimate user's and illegitimate users' data in the training set, respectively; and $\int_\beta^1 p(\varepsilon_l)d\varepsilon_l$ and $\int_\beta^1 p(\varepsilon_a)d\varepsilon_a$ denote the probabilities that the illegitimate bubble contains behavior scores derived from the legitimate user's and illegitimate users' data in the training set, respectively. If the ratio $\frac{\int_0^\alpha p(\varepsilon_l)d\varepsilon_l}{\int_0^\alpha p(\varepsilon_a)d\varepsilon_a}$ is large, it indicates that the legitimate user's behavior scores still dominate the legitimate bubble, and the distance of \emph{Privilege Movement} is appropriate. Otherwise, the distance needs to be adjusted.

%\subsection{noise filtering}
%As we discussed in previous section, human behaviors are ever changing and as shown in Fig. \ref{fig:eva} some behaviors drift away from normal behaviors which can cause false reject and false accept. To reduce the behavior drifting and filter out noise data, we use Kalman filter that discussed in Section. .....

\section{Performance Evaluation \label{pe}}
We have established a large-scale synthetic environment using MIT Friends and Family Dataset \cite{p21,aharony2011social}, and have conducted several experiments on BMap that contains a top level, two observation levels, and a bottom level. The MIT Friends and Family Dataset contains 130 participants and has a total of 9 features (GPS, accelerometer, SMS, app installation, battery usage, call logs, app usage, blue-tooth devices log, Wi-Fi access points) recorded over five months. It is a complete dataset about human behavior based on sensor data. Sensitive information such as phone numbers and chat history has been hashed to protect users' privacy. The detail of the dataset and its collecting process can be found in \cite{p21,aharony2011social}. In addition, we have implemented Shi scheme \cite{p9} and Multi-Sensor scheme \cite{p62} for comparison purposes in the simulation. To evaluate the EER, we used the recommended settings of original papers \cite{p9,p62}. We adopted k-fold cross-validation \cite{getoor2007introduction} to choose the best value of the parameters in each scheme, and to conduct training and testing. The simulation uses data from all 130 participants in five months, where we randomly selected one participant as the legitimate user and mixed her data with the data sampled from all other participants. The simulations were performed 130 times for each participant against all the other participants and averaged results were derived for each test. The decision threshold of the original scheme is set differently for each user to provide a fair comparison. In addition, the number of samples in each time window is dynamically adjusted based on the Wind Vane algorithm \cite{Yang2016}, while the time window is set to 15 seconds. We kept the illegitimate users' data portion in the range of 49\% to 51\% to construct a balanced dataset. Features contained in the dataset include GPS, accelerometer, SMS, app installation, battery usage, call logs, app usage, blue-tooth devices log, and Wi-Fi access points. A detailed description of the features can be found in the paper \cite{aharony2011social}. In addition, the feature selection strictly follows the description of the original papers \cite{p9,p62}.

\begin{figure}[htb]
\centering
  \begin{subfigure}[b]{0.5\linewidth}
  \hspace*{-0.2cm}
    \centering
    \includegraphics[width=1.9in,height=1.5in]{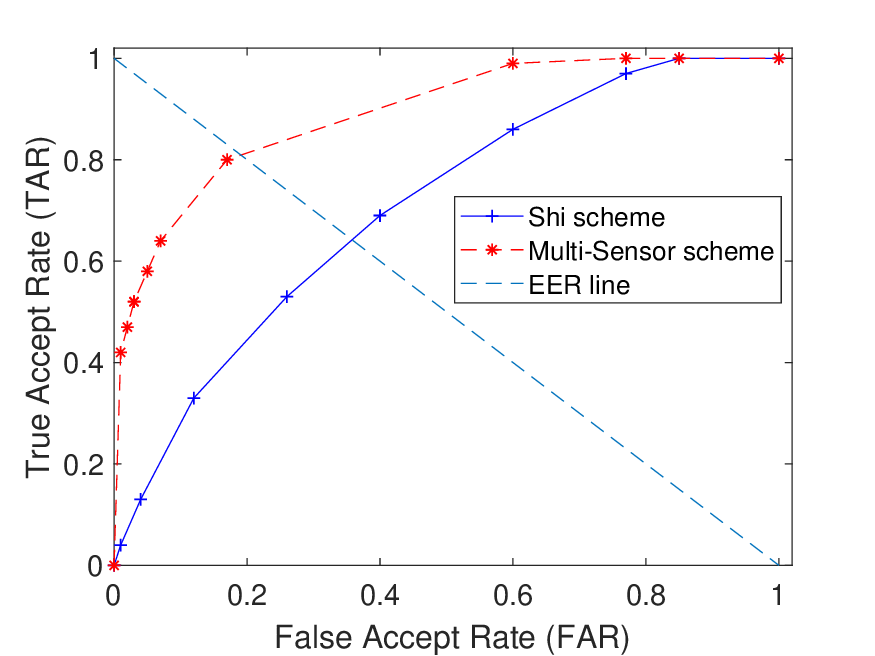}
    \caption{}\label{fig:acca}
  \end{subfigure}%
  \begin{subfigure}[b]{0.5\linewidth}
  \hspace*{-0.2cm}
    \centering
    \includegraphics[width=1.9in,height=1.5in]{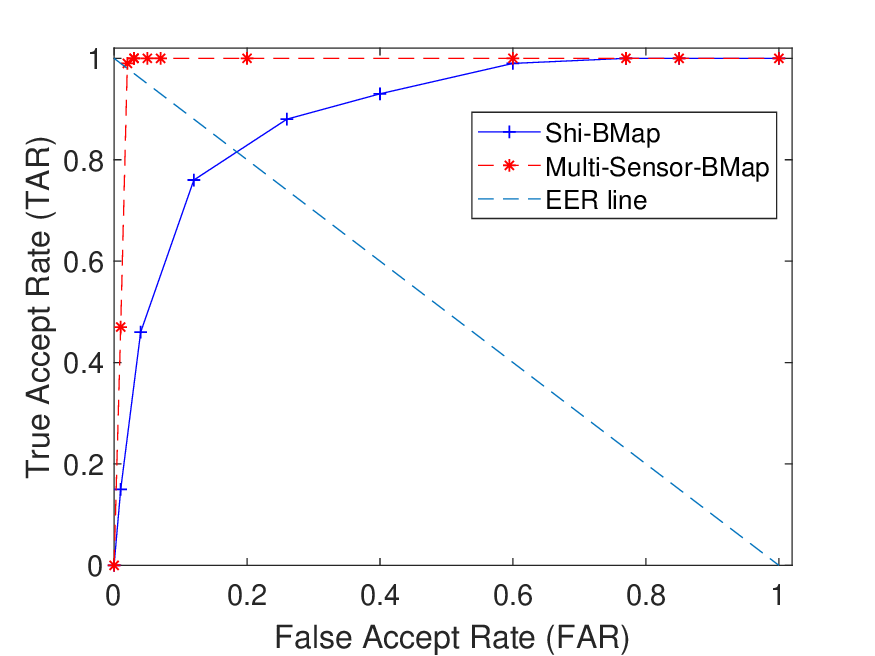}
    \caption{}\label{fig:acca}
  \end{subfigure}%
  \caption{The ROC curve of (a) original schemes and (b) BMap-based schemes.}
  \label{fig:acc}
\end{figure}

\subsection{Equal Error Rate \label{section:eer}}
The time window was set to 15 seconds in the test, which contains 1 KB of user data.  To analyze the Equal Error Rate (EER) of various schemes, we plotted the Receiver Operating Characteristic curve (ROC) using the true accept rate (TAR) and the false accept rate (FAR). To understand the trade-off between TAR and FAR, we thresholded the behavior score. Specifically, for Shi scheme, we thresholded the behavior score derived by a Gaussian mixture model. For Multi-Sensor scheme, according to the original paper \cite{p62}, we implemented it using LIB-SVM \cite{chang2011libsvm} with RBF kernel, where we thresholded both $\gamma$ and $C$ parameters. Finally, among all users, we evaluated the EER for Shi scheme and Multi-Sensor scheme. We then applied BMap to these two schemes and evaluated their corresponding EERs. The results are shown in Fig. \ref{fig:acc}. In Fig. \ref{fig:acc} (a), Shi scheme denotes an original scheme proposed by Shi et al. \cite{p9}; Multi-Sensor scheme denotes an original scheme proposed by Lee et al \cite{p62}. The corresponding schemes after applying BMap are Shi-BMap and Multi-Sensor-BMap, as shown in Fig. \ref{fig:acc} (b).

As shown in Fig. \ref{fig:acc}, the EERs for both Shi scheme and Multi-Sensor scheme are significantly reduced after BMap is applied. We further calculated the EER and area under the ROC curve (AUC) of various schemes. The EER of Shi scheme and Shi-BMap are 0.3580 and 0.1846, respectively. The EER of Multi-Sensor scheme and Multi-Sensor-BMap are 0.1908 and 0.0198, respectively. The AUC of Shi Scheme and Shi-BMap are 0.7061 and 0.8913, respectively. The AUC of Multi-Sensor and Multi-Sensor-BMap are 0.8907 and 0.9896, respectively. In addition, we calculated the statistical significance of the accuracy ($ACC=\frac{TA+TR}{TA+TR+FA+FR}$) improvement for both schemes using a paired t-test given a significance level of 0.05 with a null hypothesis: ``$H_0: $ there is no difference between the original scheme and the BMap-based scheme in terms of authentication accuracy". The result successfully rejects the null hypothesis with a $p-value=0.0159$ for Shi scheme and $1.7346\times10^{-96}$ for Multi-Sensor scheme, respectively. Hence, the accuracy improvement of BMap is statistically significant for both schemes.

%As shown in Fig. \ref{fig:acc}, the authentication accuracies for both the Shi scheme and Multi-sensor scheme are significantly improved after BMap is applied. Specifically, the authentication accuracies of Shi-BMap and Multi-Sensor-BMap increase at the beginning and then maintain high accuracies for the rest of the tests. In this dataset, the average accuracies for Shi scheme, Multi-sensor scheme, Shi-BMap, and Multi-Sensor-BMap are 64.12\%, 81.26\%, 97.44\%, and 99\%, respectively. The accuracy boost of using BMap is significant for both schemes.

\begin{figure}[htb]
  \begin{subfigure}[b]{1\linewidth}
  \hspace*{-0.5cm}
    \centering
    \includegraphics[width=3.6in,height=2.2in]{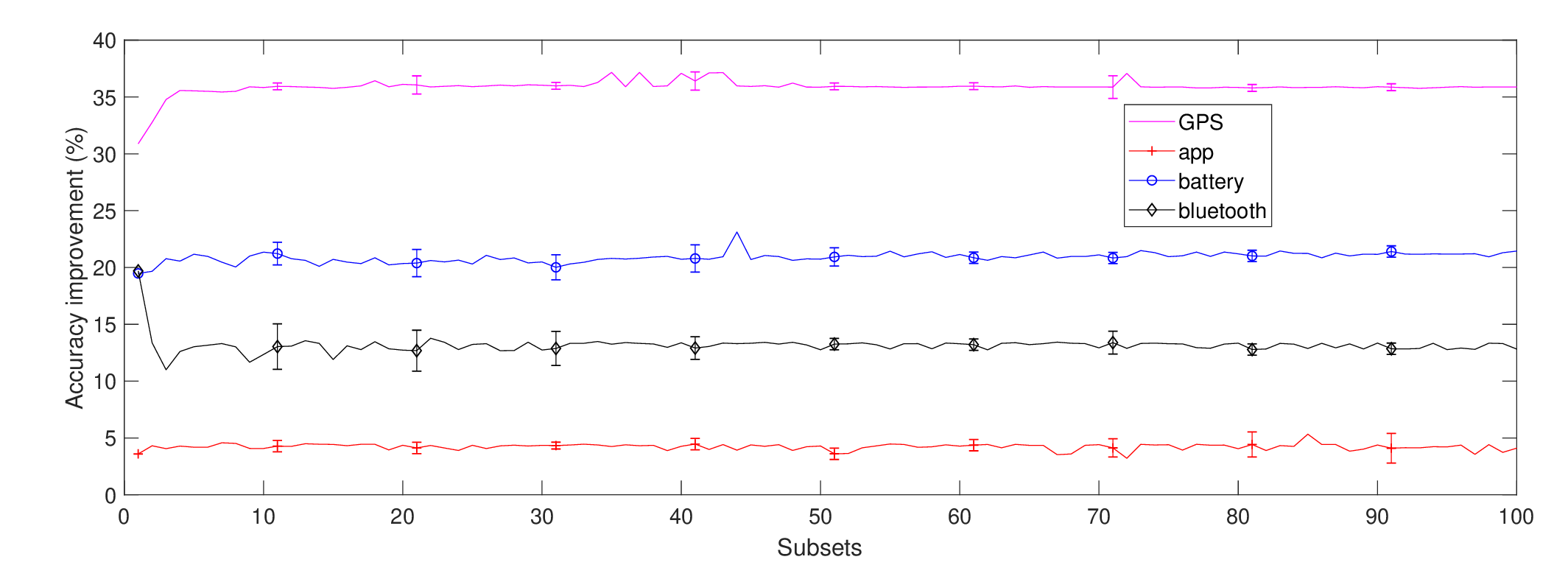}
    %\caption{}\label{fig:accb}
  \end{subfigure}%
  \caption{Accuracy improvement for individual features. \footnotesize Note that the accuracy improvement is calculated by subtracting the accuracy of the original scheme (Multi-Sensor) from the BMap boosted scheme (Multi-Sensor-BMap).}
  \label{fig:accFeatures}
\end{figure}

To simulate real usage, we divided the whole dataset into 100 distinct subsets sorted based on time, and performed tests by gradually sending the subsets to the system. We used Multi-Sensor scheme for each feature and calculated their accuracy improvement after applying BMap. In this experiment, to optimize the performance of Multi-Sensor scheme, all the parameters are finely tuned. Fig. \ref{fig:accFeatures} provides a more detailed view of accuracy improvement in GPS, app installation, battery usage, and blue-tooth devices log. Another feature has similar patterns. In Fig. \ref{fig:accFeatures}, the accuracy improvement for BMap is not stable during the first few authentications because of \emph{Bubble Expansion}. It becomes stable after the 70th subset for all four features. Generally, the accuracy improvement after applying BMap is between 4\% and 35\%.

\subsection{Performance under Large-scale Usage}

\begin{table}[!ht]
%\footnotesize
%% increase table row spacing, adjust to taste
\renewcommand{\arraystretch}{1.3}
% if using array.sty, it might be a good idea to tweak the value of
% \extrarowheight as needed to properly center the text within the cells
\caption{Performance evaluation under large-scale usage.}
%\label{table_example}
\centering
%% Some packages, such as MDW tools, offer better commands for making tables
%% than the plain LaTeX2e tabular which is used here.
\begin{center}
%\hspace{-0.47cm}
\footnotesize
\begin{tabular}{|c|c|c|c|c|c|c|}
\hline
\multicolumn{7}{ |c| }{Original Multi-Sensor scheme ($\pm 1.0$) \%} \\
\hline
Time*&ACC &PREC &TAR &TRR &FAR &FRR \\
\hline
200 &87.59&92.01&89.17&69.39&30.62&10.82\\
\hline
300 &84.40&87.77&87.08&67.84&32.15&12.93\\
\hline
500 &83.16&86.90&84.62&66.77&33.23&15.37\\
\hline
\hline
\multicolumn{7}{ |c| }{Multi-Sensor-BMap ($\pm 1.0$) \%} \\
\hline
200 &97.25&97.40&98.81&93.53&6.48&1.20\\
\hline
300 &98.64&98.87&98.93&98.18&1.82&1.08\\
\hline
500 &98.96&99.08&99.14&98.72&1.28&0.85\\
\hline
\end{tabular}
\end{center}
\begin{tablenotes}
      \footnotesize
      \item *Time stands for time window. $ACC=\frac{TA+TR}{TA+TR+FA+FR}$, $PREC=\frac{TA}{TA+FA}$, $TAR=\frac{TA}{TA+FR}$, $TRR=\frac{TR}{TR+FA}$, $FAR=\frac{FA}{FA+TR}$ and $FRR=\frac{FR}{FR+TA}$.
\end{tablenotes}
\label{tbperformance}
\end{table}

We evaluated the performance of BMap under large-scale usage using data from all users in three time slots containing 200, 300, and 500 time windows. In this experiment, instead of using all the data in the dataset, we evaluated Multi-Sensor scheme only using the data in each time slot. Specifically, we trained the SVM and tested the performance of the original scheme and BMap-based scheme in each time slot.  In this experiment, we adopted ten-fold cross-validation to tune parameters such that an optimized (smallest) EER is achieved. Other settings remain unchanged, as in Section \ref{section:eer}. We calculated the accuracy (ACC), precision (PREC), true accept rate (TAR), true reject rate (TRR), false accept rate (FAR), and false reject rate (FRR) in Table \ref{tbperformance} for both the original Multi-Sensor scheme and Multi-Sensor-BMap. As shown in Table \ref{tbperformance}, the performance improvement after applying BMap is significant compared to the original scheme. Another important observation is that the performance of the original Multi-Sensor scheme does not monotonically increase with time. In the other words, the authentication accuracy of Multi-Sensor scheme does not always improve as we gather more behavior data. This is due to behavior deviation and sensor noise. By applying BMap to the original Multi-Sensor scheme, the system becomes more predictable in terms of improving the authentication accuracy since it automatically corrects behavior deviation and filters out noise in each authentication cycle.

Furthermore, as shown in Table \ref{tbperformance}, BMap's accuracy improvement becomes smaller between the 300 and 500 time windows, compared with between the 200 and 300 time windows. As discussed previously, BMap reduces the impact of overlapping behavior scores in the slack bubble using \emph{Initial Mapping}, \emph{Privilege Movement}, and \emph{Bubble Expansion}. Since it is loop carried, the accuracy improvement is reflected gradually in each time window, and the expansion becomes slower and more stable with time. Note that we used scores that fall into any of these three bubbles to calculate the accuracy. If a behavior score derived from a legitimate user falls into the slack bubble or illegitimate bubble, it will be considered a false reject; otherwise, it will be considered a true accept. If a behavior score derived from an illegitimate user falls into the slack bubble or legitimate bubble, it will be considered a false accept; otherwise, it will be considered a true reject.

In addition, using Multi-Sensor-BMap, we also simulated password input for both legitimate users and illegitimate users. Specifically, we assume 94\% of legitimate users correctly input the password in each try based on the survey result shown in \cite{p14}. We also assume it takes illegitimate users at least three tries to successfully guess the correct password. In practice, however, the average number of guesses needed to pass the authentication is much larger than three for most systems \cite{bellovin1992encrypted,gong1995optimal}. In the test, there is no illegitimate user being mapped to the top privilege level; and 98\% of legitimate users have been mapped to the top level since the beginning of usage.

\section{Implementation}
In the previous section, we mainly focus on the short-term evaluation under a large-scale synthetic environment. In practice, a long-term evaluation is also important. In addition to the evaluations of Shi-BMap and Multi-Sensor-BMap, we want to measure the performance of BMap on other state-of-the-art implicit authentication schemes, e.g., Gait scheme \cite{frank2010activity}, SilentSense scheme \cite{bo2013silentsense}, and Touchalytics scheme \cite{frank2013touchalytics}. To measure the performance of these schemes in real usage requires us to implement several back-end services and servers for training and testing purposes. To this end, we implemented a BMap-based system using Android smartphones and multiple servers. We developed user-side services to achieve data sampling, data storing, data packaging, noise filtering, and authentication. The sampling algorithm is achieved using the wind-vane framework \cite{yang2017energy}. We implemented a database server to store users' data and further filtered noise and invalid data samples. We also implemented another server, which is independent of the database server for training purposes.

\section{Real Experiment}

\begin{table}[!ht]
\footnotesize
%% increase table row spacing, adjust to taste
\renewcommand{\arraystretch}{1.3}
% if using array.sty, it might be a good idea to tweak the value of
% \extrarowheight as needed to properly center the text within the cells
\caption{Demographic background}
%\label{table_example}
\centering
%% Some packages, such as MDW tools, offer better commands for making tables
%% than the plain LaTeX2e tabular which is used here.
\begin{center}
\hspace*{-0.4cm}
\begin{tabular}{|c|c|c|c|c|c|}
\hline
    &Student&Faculty&Non-faculty*&M.&F.\\
\hline
Age21-30&7&0&1&6&2\\
\hline
Age31-40&2&0&1&2&1\\
\hline
Age41-50&0&1&1&1&1\\
\hline

%Batt(\%)&0.6&2.8&5.3&0.2&1.0\\
%\hline
\hline
\end{tabular}
\end{center}
\begin{tablenotes}
      \footnotesize
      \item * Non-faculty contains one who is not employed by the university. M. means male. F. means female.
\end{tablenotes}
\label{tbDB}
\end{table}

To evaluate the performance of BMap, we conducted a long-term real test from 2016 to 2020. We recruited students and faculty from the University of Tennessee as volunteers for this experiment. Over four years, we analyzed the behavioral data of 13 different volunteers using the proposed system. In order to ensure user privacy, in the experiment, we did not meet with any participants who picked up their devices from our lab on their own. We used time stamps and device ids stored in Google Firebase to uniquely identify each user. The demographic information is shown in Table \ref{tbDB}, which contains occupation (student, faculty, or non-faculty), age, gender, and education. We used a total of three Android devices (Samsung Nexus S, Samsung Galaxy S10, and Motorola G2). Each device only had one legitimate user at one testing period, while other users were deemed illegitimate. It is possible for one user to be a legitimate user for one device and an illegitimate user for another device at the same time. Every participant, in turn, was selected as a legitimate user and continuously used one of the devices for at least two weeks and, on average, eleven months in total. The device was reset, and the data was cleared before being handed to the next legitimate user. We did not restrict users to perform any particular operation using the device, but they were required to use the device at least four hours per day. In addition, every illegitimate user was encouraged to guess the password and mimic legitimate users' behavior during usage. Every two weeks, every participant would be gathered in the lab by an administrator who does not belong to this project. The legitimate user who completed their test would hand the device to the next legitimate user. Otherwise, the legitimate user would give the device to one of the illegitimate users who had not used their device before and was willing to spend at least half to one hour to observe the legitimate user's behavior. It is possible that the legitimate user is a close friend of the illegitimate user. If a mimicry attacker could not be found, the device would be given to one of the illegitimate users who did not take the device before. At the end of the test, it is guaranteed that for each legitimate user, there is at least one illegitimate user who mimicked the legitimate user's behavior; and other illegitimate users all used the device. The illegitimate user would occupy this device for at least two weeks to guess the password and mimic the legitimate user's behavior before handing back the device to the legitimate user. The dataset used in the experiment was adjusted to contain both the legitimate (50\%) and illegitimate (50\%) users' data, where the mimicry data was kept above 60\% of the illegitimate users' data.

%To this end, the devices' passwords are randomly chosen with a length of eight characters that contain both letters and numbers.

%For each device, one of 13 volunteers was using it in a period of time (longer than two weeks); and thus, each device stores data from 12 illegitimate users and one legitimate user.

In the previous experiment, due to the limitation of the dataset, we can only evaluate BMap on Shi scheme and Multi-Sensor scheme. However, in the long-term real test, we gathered rich usage information from all users. Besides the features used in the synthetic experiments, we also collected touch-related data, e.g., trajectory, pressing time, and corresponding accelerometer reading, which makes the evaluation of BMap on Gait scheme \cite{frank2010activity}, SilentSense scheme \cite{bo2013silentsense}, and Touchalytics \cite{frank2013touchalytics} possible. To this end, we implemented Gait scheme, SilentSense scheme, and Touchalytics scheme in our system. Similar to the implementation of Shi scheme and Multi-Sensor scheme, we used the recommended settings of Gait scheme, SilentSense scheme, and Touchalytics scheme from their original papers. In addition, we applied k-fold cross-validation to choose the best value of the parameters in each scheme and conduct training and testing. The feature selection strictly follows the description of the original papers. Meanwhile, the decision threshold of the original scheme is set differently for each user to provide a fair comparison.

\subsection{Equal Error Rate}

\begin{figure}[htb]
\centering
  \begin{subfigure}[b]{0.5\linewidth}
  \hspace*{-0.2cm}
    \centering
    \includegraphics[width=1.9in,height=1.5in]{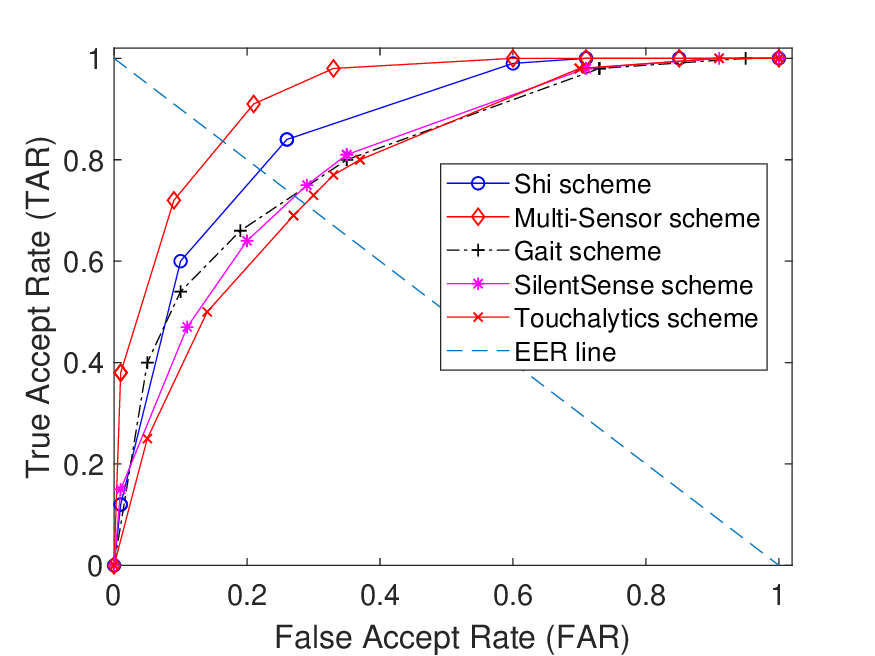}
    \caption{}\label{fig:acca}
  \end{subfigure}%
  \begin{subfigure}[b]{0.5\linewidth}
  \hspace*{-0.2cm}
    \centering
    \includegraphics[width=1.9in,height=1.5in]{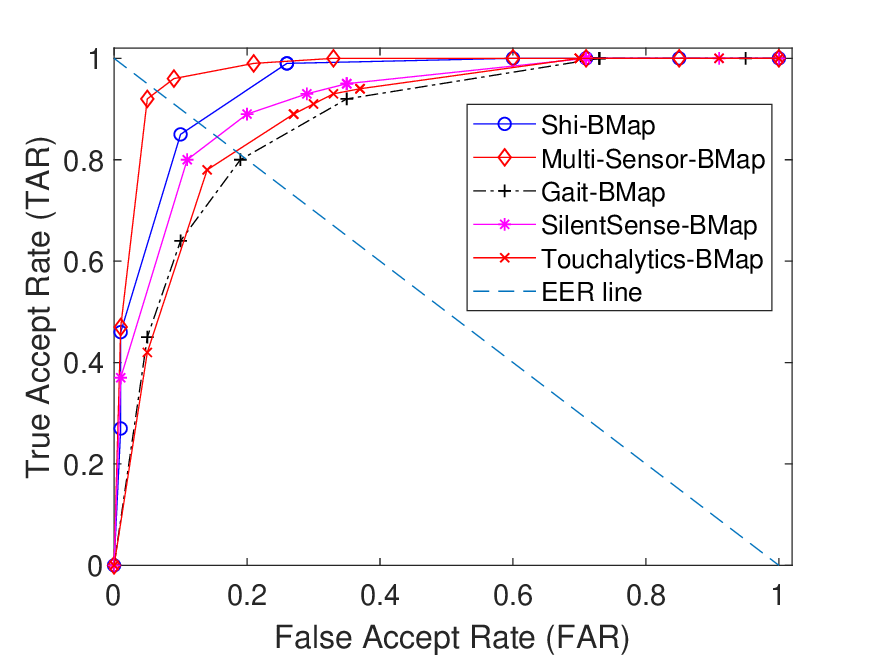}
    \caption{}\label{fig:acca}
  \end{subfigure}%
  \caption{The ROC curve of (a) original schemes and (b) BMap-based schemes.}
  \label{fig:accInRealTest}
\end{figure}

To evaluate the EERs' improvement of BMap on different schemes, we plotted the ROC curve using the TAR and FAR. To understand the trade-off between TAR and FAR, we thresholded the behavior score. Specifically, for Shi scheme, we thresholded the behavior score derived by a Gaussian mixture model. For Multi-Sensor scheme, SilentSense scheme, and Touchalytics scheme we implemented them using SVM with RBF kernel, where we thresholded both $\gamma$ and $C$ parameters. For Gait scheme, we thresholded the similarity value of the k nearest neighbor. We then measured the EER and AUC of original schemes using the testing dataset. Under the same setting, we then applied BMap to the schemes, and repeated the measurement on the same testing dataset. The results are shown in Fig. \ref{fig:accInRealTest}. As shown in the figure, BMap-based schemes have lower EERs than all the original schemes. We calculated the EERs for original Shi scheme, Muti-Sensor scheme, Gait scheme, SilentSense scheme, and Touchalytics scheme, which are 0.2200, 0.1635, 0.2700, 0.2720, and 0.2871 respectively; and corresponding AUCs are 0.8588, 0.9244, 0.8103, 0.8013, and 0.7842,respectively. We also calculated the EERs for BMap-based Shi scheme, Multi-Sensor scheme, Gait scheme, SilentSense scheme, and Touchalytics scheme, which are 0.1267, 0.0650, 0.1957, 0.1550, and 0.1833, respectively; and corresponding AUCs are 0.9458, 0.9741, 0.8757, 0.9157, and 0.8851, respectively. For all four schemes, BMap boosts their performance significantly, especially for SilentSense scheme. The EER improvements of Shi scheme, Muti-Sensor scheme, Gait scheme, SilentSense scheme, and Touchalytics schemes are 0.0933, 0.0985, 0.0743, 0.1170, and 0.1038, respectively. Although none of the illegitimate users successfully guessed the correct passwords, they can still mimic legitimate users' behavior and pass the authentication, which is one of the biggest problems in today's implicit authentication schemes \cite{khan2016targeted}. Especially for Gait scheme, SilentSense, and Touchalytics scheme, their corresponding mimicry attacks are very effective. However, as shown in Fig. \ref{fig:accInRealTest}, BMap can still reduce the success rate of the attacks and EERs of the schemes.

As discussed in Section \ref{da}, the adversary can brute force the passwords, but it also expands the illegitimate bubble to cover all the whole interval ($[0$ $1]$) and causes an immediate locking of the device after inputting the passwords. Although adversary can observe legitimate user's behavior without touching the device, to fully mimic the behavior and launch the attack, it requires multiple attempts \cite{khan2016targeted}. We recorded the number of tries for the adversary to successfully mimic legitimate user's behavior; and the result shows at least five attempts (2.5 minutes on average) are needed to pass the authentication, which agrees with the result in \cite{khan2016targeted}. On another aspect, besides the observation, it also requires the adversary to spend time using the device to pass the authentication. Both observation and using time contribute to the launch time of the mimicry attack. In the end, due to \emph{Privilege Movement}, even though the attacker can perfectly mimic legitimate users' behavior, they will be blocked by the device.

\begin{table}[!ht]
\footnotesize
%% increase table row spacing, adjust to taste
\renewcommand{\arraystretch}{1.3}
% if using array.sty, it might be a good idea to tweak the value of
% \extrarowheight as needed to properly center the text within the cells
\caption{The number of times users been locked out}
%\label{table_example}
\centering
%% Some packages, such as MDW tools, offer better commands for making tables
%% than the plain LaTeX2e tabular which is used here.
\begin{center}
\hspace*{-0.4cm}
\begin{tabular}{|c|c|c|c|c|}
\hline
        Scheme& \multicolumn{2}{|c|}{Original scheme} & \multicolumn{2}{|c|}{BMap-based scheme}\\
        \hline
\hline
    &Legi.* &Ille.* &Legi.* &Ille.* \\
\hline
Shi scheme &213&4863&179&6243\\
\hline
MultiSensor &187&5640&53&7015\\
\hline
Gait scheme &379&4310&205&5186\\
\hline
Silent Sense &179&3973&174&5800\\
\hline
Touchalytics &189&3847&183&5389\\
\hline

%Batt(\%)&0.6&2.8&5.3&0.2&1.0\\
%\hline
\hline
\end{tabular}
\end{center}
\begin{tablenotes}
      \footnotesize
      \item * Legi. denotes the legitimate user. Ille. denotes the illegitimate user.
\end{tablenotes}
\label{tbLockTimes}
\end{table}

In this test, for both original schemes and BMap-based schemes, we stored the number of times of a legitimate user being locked out of the device within 7,250 attempts. To compare, we recorded the number of times of an illegitimate user being locked out of the device in 7,250 attempts. The testing results are shown in Table \ref{tbLockTimes}. Comparing to the original scheme and BMap-based scheme, we can see the number of times a legitimate user has been locked out is reduced; and hence, the usability of the system is increased. Similarly, the number of times an illegitimate user has been locked out is increased. The security of the system is enhanced since an illegitimate user will have a higher chance of being blocked.

\subsection{Time Consumption}

We evaluated the time consumption of BMap and compared it with the time consumption of data transmitting, data initialization, and data exporting in the original systems. Table \ref{tbTimeConsumption} shows the time consumption for different schemes. The column denotes different stages of various schemes. As shown in the table, the time consumption of BMap on different schemes is very small compared to other operations. The total time consumption of each scheme after applying BMap is shown in the final column of the table.

%As shown in Table \ref{tbTimeConsumption}, MultiSensor has the highest time consumption among all four schemes, which takes a total of 52.66 seconds on average. Since MultiSensor selects more feature than another scheme, it significantly increases the training time and transmission time. Utilizing a small amount of feature, Shi scheme has the lowest time consumption among all the schemes. Although the difference of the total time consumption between MultiSensor and Shi scheme is large (48.53 seconds), on BMap part the difference is very small (less than 0.02 second). To achieve high authentication accuracy and large coverage in practice, existing IA systems tend to choose more and more features to identify users. Since BMap is not sensitive to the size of the feature set, it is naturally suitable for these systems.

\begin{table}[!ht]
\footnotesize
%% increase table row spacing, adjust to taste
\renewcommand{\arraystretch}{1.3}
% if using array.sty, it might be a good idea to tweak the value of
% \extrarowheight as needed to properly center the text within the cells
\caption{Time consumption for different schemes (sec)}
%\label{table_example}
\centering
%% Some packages, such as MDW tools, offer better commands for making tables
%% than the plain LaTeX2e tabular which is used here.
\begin{center}
%\hspace*{-0.4cm}
\footnotesize
\begin{tabular}{|c|c|c|c|c|c|}
\hline
    &BMap&Trans.*&Init.*&Export&Total\\
\hline
Shi scheme &0.046&0.96&1.57&1.03&3.61\\
\hline
MultiSensor &0.051&4.69&2.137&5.16&12.04\\
\hline
Gait scheme &0.127&0.95&2.917&1.03&5.02\\
\hline
SilentSense &0.140&1.03&1.918&1.06&4.15\\
\hline
Touchalytics &0.136&0.95&2.703&1.02&4.81\\
\hline

%Batt(\%)&0.6&2.8&5.3&0.2&1.0\\
%\hline
\hline
\end{tabular}
\end{center}
\begin{tablenotes}
      \footnotesize
      \item * Trans. denotes the total data transmission time consumption in the system except BMap. Init. denotes the time consumption of data initialization. In addition, the data initialization contains data formatting and noise filtering.
\end{tablenotes}
\label{tbTimeConsumption}
\end{table}

\begin{figure}[htb]
\centering
  \begin{subfigure}[b]{.5\linewidth}
  %\hspace*{-0.2cm}
    \centering
    \includegraphics[width=1.099\textwidth,height=1.4in]{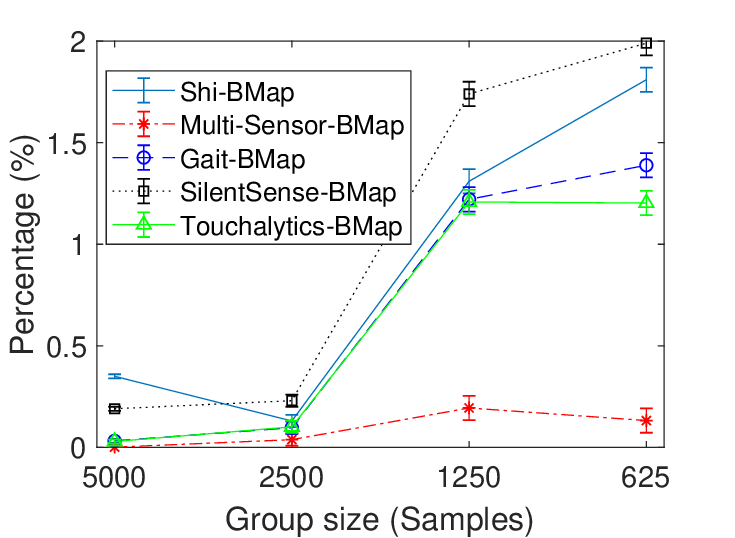}
    \caption{}\label{fig:tcrta}
  \end{subfigure}%
  \begin{subfigure}[b]{.5\linewidth}
  %\hspace*{0.1cm}
    \centering
    \includegraphics[width=1.1\textwidth,height=1.4in]{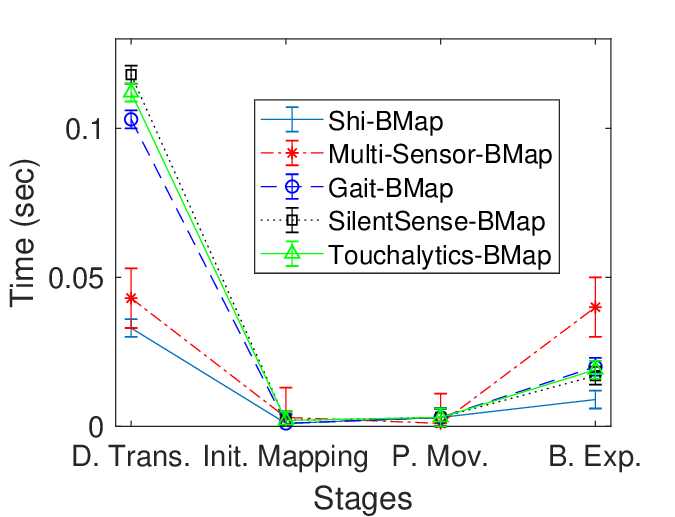}
    \caption{}\label{fig:tcrtb}
  \end{subfigure}%
  \caption{Time Consumption. (a) Time consumption percentage in different group sizes. (b) Time consumptions of different stages.}
  \label{fig:timeConsumptionRealTest}
\end{figure}

In addition, since implicit authentication utilizes a group of data exported recently to identify users, e.g., 5,000 samples in each group, the size of the group impacts the time consumption of the system. Given different data exporting frequency, the time increment of BMap is shown in Fig. \ref{fig:timeConsumptionRealTest} (a). Furthermore, among different group sizes, we calculated the average time-consumption percentages of BMap in different schemes, which are 0.9\%, 0.0912\%, 0.657\%, 1.037\%, and 0.6353\% for Shi scheme, MultiSensor scheme, Gait scheme, SilentSense scheme, and Touchalytics scheme correspondingly. Specifically, the time consumptions of BMap in Data Transmission, \emph{Initial Mapping}, \emph{Privilege Movement}, and \emph{Bubble Expansion} are shown in Fig. \ref{fig:timeConsumptionRealTest} (b).

%In Fig. \ref{fig:timeConsumptionRealTest} (a), Multi-Sensor-BMap has a stable time consumption percentage among all four groups. Although the time consumption percentages for SilentSense, Shi, and Gait schemes are not as stable as MultiSensor scheme, the fluctuation is very small (less than $\pm 1.8\%$). In addition, given large group size, the differences among the schemes are small (less than $\pm 0.3\%$). As shown in Fig. \ref{fig:timeConsumptionRealTest} (b), Multi-Sensor-BMap has the highest time consumption at all four stages, but the difference between the highest and lowest time consumption is small (less than $\pm 0.1\%$). The time consumptions for various schemes have some small differences, especially at \emph{Initial Mapping} and \emph{Privilege Movement} stages.

\subsection{Energy Consumption}

\begin{figure}[htb]
\centering
  \begin{subfigure}[b]{.5\linewidth}
  \hspace*{-0.2cm}
    \centering
    \includegraphics[width=1.099\textwidth,height=1.4in]{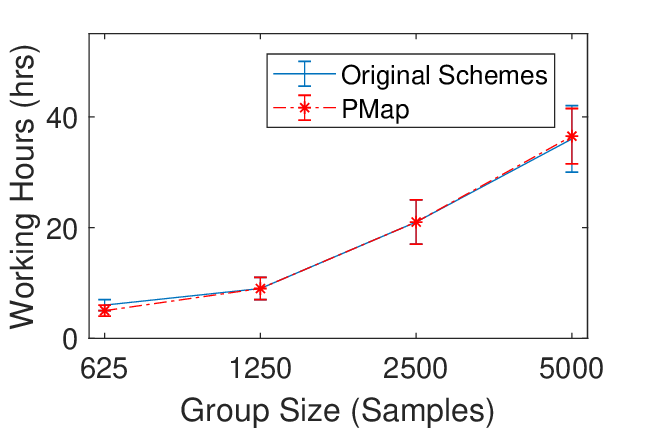}
    \caption{}\label{fig:ecb}
  \end{subfigure}%
  \begin{subfigure}[b]{.5\linewidth}
  %\hspace*{0.1cm}
    \centering
    \includegraphics[width=1.099\textwidth,height=1.4in]{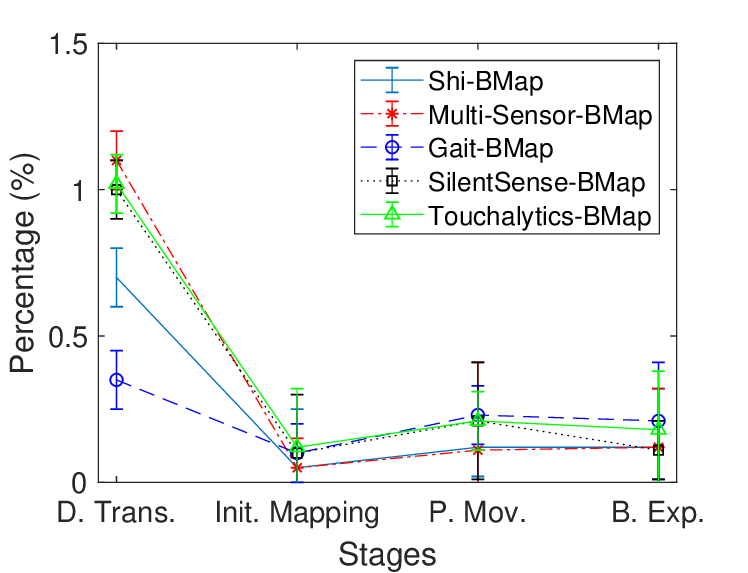}
    \caption{}\label{fig:eca}
  \end{subfigure}%
  \caption{Battery Consumption. (a) The average battery consumption for both original schemes and BMap-based schemes. (b) The average battery consumption of different stages.}
  \label{fig:energyConsumptionFigure}
\end{figure}

Besides the time consumption, we also conducted several experiments to measure the difference between original schemes and BMap-based schemes in the aspect of energy consumption. In the experiment, we measured the battery usage in the original schemes by calculating the average working hours of the battery after fully charged. To compare, we also measured the battery usage in the BMap-based schemes. The details are shown in Fig. \ref{fig:energyConsumptionFigure}. As shown in Fig. \ref{fig:energyConsumptionFigure} (a), the average working hours of original schemes and BMap-based schemes are almost the same. Specifically, we calculated the battery working hour reduction by applying BMap, which is less than 0.9\% of the total working time. The average battery consumption in Data Transmission, \emph{Initial Mapping, Privilege Movement, and Bubble Expansion} is shown in Fig. \ref{fig:energyConsumptionFigure} (b). It is calculated using the default system tool in Android. In the experiment, since we utilized a secure channel to transmit users' data, it consumes the most portion of energy, especially for Multi-Sensor-BMap. In addition, the private data hash and formatting are conducted in this stage. In the rest of stages, \emph{Initial Mapping, Privilege Movement, and Bubble expansion} have similar energy consumption.

\vspace*{-0.4cm}
\subsection{Other Performance Measures}

\begin{figure}[htb]
\vspace*{-0.5cm}
\centering
  \begin{subfigure}[b]{.5\linewidth}
  \hspace*{-0.2cm}
    \centering
    \includegraphics[width=1.099\textwidth,height=1.4in]{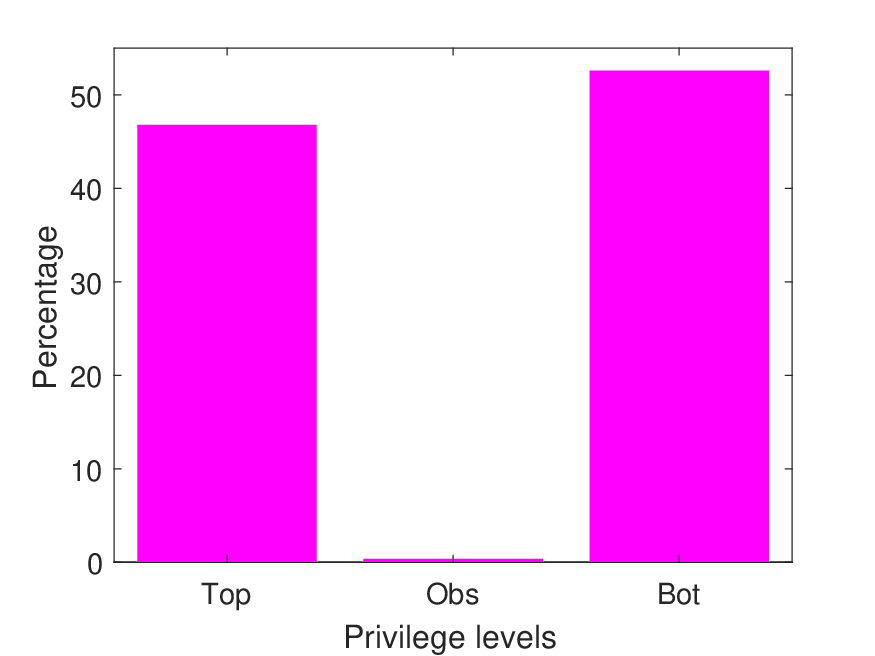}
    \caption{}\label{fig:lba}
  \end{subfigure}%
  \begin{subfigure}[b]{.5\linewidth}
  %\hspace*{0.1cm}
    \centering
    \includegraphics[width=1.099\textwidth,height=1.4in]{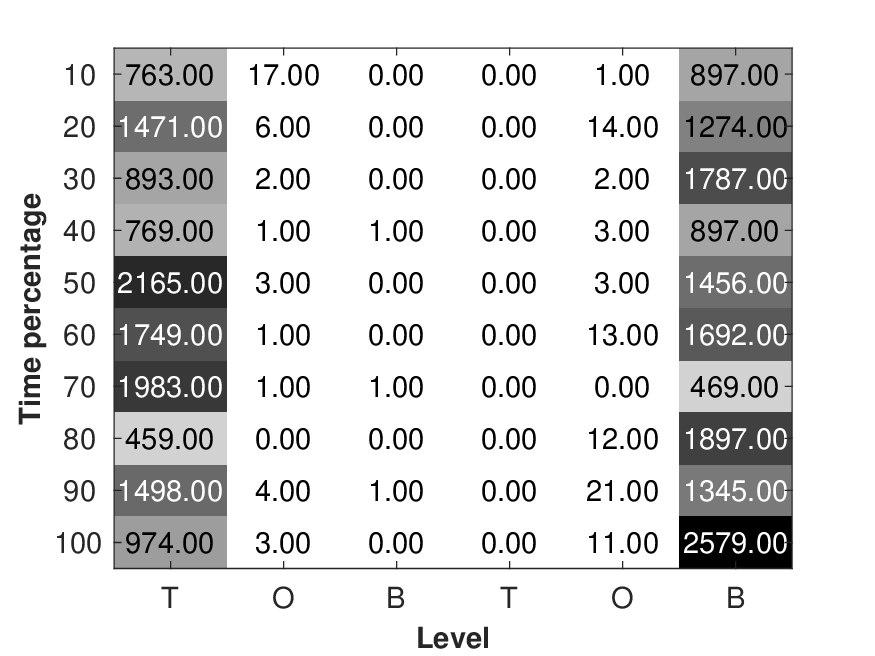}
    \caption{}\label{fig:lbb}
  \end{subfigure}%
  \caption{The proportion of behavior scores in each level. (a) Average proportion. (b) Proportion across one month.}
  \label{fig:levelbin}
\end{figure}

By using the data sampled from 2016 to 2019, we calculated the percentage of behavior scores that were mapped to each privilege level in BMap-based Multi-Sensor scheme, for the legitimate user and illegitimate users. As shown in Fig. \ref{fig:levelbin} (a), less than 0.4\% of the behavior scores are mapped to the observation levels. We also calculated the average time users spend at the observation level, which is only 57 seconds per day. The result indicates that BMap is fast and highly effective in making the final decision.

%As discussed in the previous sections, the observation levels are used to monitor the current user's behavior before making the decision. The optimal situation is to immediately map the legitimate user to top privilege level and the illegitimate user to bottom level, but in practice, due to the behavior changes it is very hard to achieve immediate mapping without decreasing authentication accuracy. Spending a small amount of time to observe users' behavior, BMap universally improves the authentication accuracy of the IA system.

In order to analyze the detailed change of each level, we magnified the experimental result to check the behavior score distributions at each privilege level during one month's usage. The result is shown in Fig. \ref{fig:levelbin} (b), where we calculated the number of scores that fall into each level for all 13 participants. The x-axis denotes the privilege levels, where the left three levels that contain top, observation, and bottom levels, are plotted from the legitimate users' behavior scores. Similarly, the right three levels are plotted from illegitimate users. The y-axis indicates the time percentage, from 10\% to 100\% of the given month. The digit in the matrix denotes the number of behavior scores. The average time the legitimate user spends at the observation and bottom level are 43 and 3 seconds per day, respectively. The average time the illegitimate user spends at the observation and top level are 80 and 0 second(s) per day, respectively. For both users, the number of scores that fall in the observation level is small, less than 0.43\%, which is similar to the result in Fig. \ref{fig:levelbin} (a).

%We also calculated the behavior score distributions in each privilege level for time windows 10 through 100 as shown in Fig. \ref{fig:levelbin} (b). The z-axis denotes the number of behavior scores. The y-axis denotes the time windows. The x-axis denotes the privilege levels, where the left three levels that contain top, observation, and bottom levels, are plotted from the legitimate users' behavior scores; similarly, the right three levels are plotted from illegitimate users. For both users, the number of scores which fall in the observation level is small, less than 118 given a total of 27,138 scores, which is similar to the result in Fig. \ref{fig:levelbin} (a).

%As shown in Fig. \ref{fig:levelbin} (b), the number of behavior scores per time window is not stable, which from another aspect demonstrates the high complexity of users' behavior. Furthermore, the number of behavior scores which falls in the illegitimate users' observation level is 5\% greater than the one which falls in the legitimate users' observation level. In other words, the system spends more time on observing the illegitimate users' behavior than the legitimate users' behavior.

\section{Related Work}
The majority of the existing implicit authentication schemes \cite{p62,p7,bo2013silentsense,feng2014tips,shi2011senguard,shahzad2013secure,castelluccia2017towards,frank2013touchalytics} focus on finding suitable behavioral features such as touch, typing, and other motions that uniquely identify users. The amount of data gathered by various sensors directly affects the accuracy of implicit authentication systems \cite{p62,yang2016personaia,p9}. By increasing the time spent in collecting users' behavior data, the accuracy of implicit authentication can be improved \cite{p62,p9} with the cost of usability. In this paper, we proposed the BubbleMap (BMap) framework to improve the authentication, accuracy, and usability of the original schemes at the same time. Dynamically adjusting privilege structure and mitigating the impact of various noises, BMap adds another layer of protection to implicit authentication systems, and is generally suitable for various IA schemes such as \cite{p9,shi2011senguard,p62,frank2010activity,ravi2005activity,bo2013silentsense,frank2013touchalytics}. BMap can also be applied to wearable devices, such as \cite{ekiz2019your, vhaduri2019multi}, to enhance their authentication accuracy.

To complement primary authentication mechanisms such as PIN and passlocks, various implicit authentication schemes have been proposed as secondary authentication mechanisms \cite{maiorana2011keystroke,p32,feng2014tips,shi2011senguard,p62,frank2010activity,bo2013silentsense,jain2004introduction,mantyjarvi2005identifying,muaaz2013analysis,riva2012progressive}. Among them, leveraging different features, Shi scheme \cite{p9}, Multi-Sensor scheme \cite{p62}, Gait scheme \cite{frank2010activity}, SilentSense scheme \cite{bo2013silentsense}, and Touchalytics \cite{frank2013touchalytics} are five different schemes that represent five research directions of state-of-the-art implicit authentications \cite{khan2014comparative,serwadda2013verifiers}. In addition, current implicit authentication research tends to adopt all the available features to achieve a better authentication accuracy \cite{p62,yang2016personaia,yang2017energy}. To evaluate the performance of BMap, we implemented Shi scheme, Multi-Sensor scheme, Gait scheme, SilentSense scheme, and Touchalytics scheme. We also show BMap can seamlessly cooperate with another framework such as \cite{yang2017energy,yang2016personaia} to improve the system's performance.

BMap utilizes privilege control to dynamically adjust users' privilege. Privilege control mechanism has been widely used in different areas to enhance systems' security\cite{sandhu1993lattice, ferraiolo2001proposed, yi2001security,hayashi2012goldilocks}. Analyzed users' data, Eiji Hayashi et al \cite{hayashi2012goldilocks} suggest to use multi-level authentication to improve the accuracy and usability of biometric-based authentication systems such as implicit authentication. Some research works implemented multi-level authentication in implicit authentication using fixed levels \cite{riva2012progressive}, which is different from the approaches adopted in this work. Implicit authentication mainly utilizes biometric behavior such as touch, motion, shake, and armswing, to identify users \cite{p66,gascon2014continuous,p23,p65,p35,kate2017authentication}. Since users' behaviors have large divergence and contain various noises \cite{yang2016personaia,p67,bo2013silentsense}, directly applying multi-level authentication to implicit authentication systems is not feasible. To this end, we analyzed the functionality of implicit authentication, mathematically modeled the privilege changing process in implicit authentication, and bridged a fine-grained privilege control to implicit authentication systems using BMap. In order to adopt sophisticated human behaviors, we upgraded the traditional fixed-level privilege control \cite{sandhu1993lattice, ferraiolo2001proposed, yi2001security,hayashi2012goldilocks,hayashi2012goldilocks,babu2015prevention,crampton2010towards,sinclair2008preventative} to support any number of privilege levels.

To deal with the behavior and sensor noises, most of the existing implicit authentication schemes use simple approaches such as resampling \cite{p62}, averaging the results \cite{bo2013silentsense,frank2013touchalytics}, or no approach at all \cite{p65,p35,kate2017authentication}. Such noises will degrade system performance in terms of authentication accuracy. The problem will be exacerbated as the size of the behavior data grows. We applied a Kalman filter \cite{welch1995introduction} to correct behavior deviation and filter out sensor noise during the authentication. We showed that a Kalman filter is naturally suitable for implicit authentication and can be implemented in practice to further improve authentication accuracy while reducing the system's latency.

\section{Discussion}
This section aims to answer and discuss some important questions about BMap.

\textbf{Bubble Expansion v.s. Retraining:} Since in BMap, \emph{Bubble Expansion} is used to tune the parameters to best match the current users' behavioral data, it can be replaced by retraining the model. We have another paper specifically focused on the model retraining problem \cite{p67}. Essentially, retraining requires users to upload their recently sampled data to a remote server, where the system will tune the parameters of the model to match the current users' behavioral patterns. In the retraining, all the parameters will be tuned and optimized to best separate different users, which significantly increases the authentication accuracy. However, retraining may expose users' sensitive data to the public due to the data uploading process. It increases the chance of various attacks. Another disadvantage of retraining is energy and time consumption since users' behavioral data often reaches megabytes which may take extra time and power to upload to the remote server. \emph{Bubble Expansion} does not require data uploading and works in the background. Since \emph{Bubble Expansion} can only tune a part of the parameters related to the thresholds, we believe that the authentication accuracy boost of \emph{Bubble Expansion} should be lower than retraining. To compare \emph{Bubble Expansion} and retraining, we need to find a suitable retraining method that can seamlessly work with BMap, e.g., to retrain the model in a fixed interval or using JS divergence to optimize the retraining frequency \cite{p67}. This could be an interesting research direction for future study.

\textbf{Users' feedback when being blocked:} When IA fails to authenticate legitimate users, it will block them from further accessing the device. Although BMap can reduce the false rejects, it cannot completely prevent it from happening. However, instead of directly locking the device, BMap will temporarily map the current user to observation levels, in which users need to input passwords if they want to gain access to the higher privilege level. By using BMap-based Multi-Sensor scheme, we calculated the percentage of time that the legitimate users were asked to input passwords when they attempted to access the higher privilege level, which is on average 1\% for all 13 participants. Given a large number of attempts, 1\% can still affect usability. But we believe being mapped to the observation level as an intermediate step is more user-friendly than being completely locked out. We did not collect the users' experience when they were mapped to the observation level. More experiments will be conducted in the future to evaluate the usability of BMap.

\textbf{The collected dataset and its impact:} The availability of suitable datasets for the comprehensive evaluation of implicit authentication systems is limited \cite{pisani2019adaptive}. To give a fair comparison between various IA schemes, a dataset containing multiple samples from different sensors in one sampling cycle is needed, which makes such datasets intrinsically rare. To the best of our knowledge, there is no publicly available dataset recently published that can be used to compare all schemes mentioned in this research. The dataset we collected aims to compare the performance enhancement of various BMap-based IA schemes, but it can also be used as a benchmark for original schemes comparison, making it invaluable. We have already decided to make our dataset publicly available in the future to benefit the research in this area, which should be another contribution of this work.

\section{Conclusion and Future Work}
In this paper, we proposed BubbleMap (BMap) to enhance the performance of various implicit authentication (IA) schemes. As a seamless overlay framework, BMap can be used to boost the performance of the original schemes. In BMap, we modeled the privilege changing process of users and bridged the privilege control mechanism to implicit authentication. To this end, we introduced \emph{Initial Mapping}, \emph{Privilege Movement}, and \emph{Bubble Expansion} techniques. In addition, we evaluated BMap in a large-scale simulation on state-of-the-art IA schemes. We also implemented BMap and performed a long-term test over four years. The test results show BMap can increase the performance of the original schemes with a small amount of energy consumption. Specifically, in the real experiment, the EERs of Shi scheme, Multi-sensor scheme, Gait scheme, SilentSense scheme, and Touchalytics scheme are 0.2200, 0.1635, 0.2700, 0.2720, and 0.2871, respectively; and the EERs after applied BMap are 0.1267, 0.0650, 0.1957, 0.1550, and 0.1833, respectively. The time consumption increased by BMap is less than or equal to 1\% for all four of the schemes. Similarly, the battery consumption increased by BMap is less than 0.9\% of the total working time. We only collected 13 participants' behavioral data, which may affect the quality of the evaluation. As part of our future work, we plan to scale up the experiment to include more participants and share the source code, parameter setting, and dataset on our website \cite{BMapURL} to benefit related research.

\section{Acknowledgement}
This work was partially supported by the US National Science Foundation (NSF) under grant CNS-1422665 and the Army Research Office (ARO) under grant 66270-CS.

\bibliographystyle{IEEEtran}
\bibliography{reference}


\end{document}